\pdfoutput=1
\documentclass[reqno,16pt]{amsart}
\usepackage[colorlinks=true]{hyperref}
\usepackage{graphicx,amsmath,amsbsy,mathtools,amssymb,amsthm,amscd,amsfonts,amsxtra}
\usepackage{graphicx}
\usepackage[skip=2pt]{caption}
\usepackage{subcaption}
\usepackage{float}
\usepackage{epsfig}
\usepackage{epstopdf}
\usepackage[a4paper, total={6in, 8in}]{geometry}

\usepackage{verbatim}

\numberwithin{equation}{section}
{\bf}{\it}
{\bf}{\it}
{\bf}{\it}
{\bf}{\it}
{\bf}{\it}
{\bf}{\it}
{\bf}{\it}

\def\beq{\begin{equation}}
\def\eeq{\end{equation}}

\def\eg/{e.g.}
\def\ie/{i.e.}

\numberwithin{equation}{section}

{\bf}{\em}
{\bf}{\em}
\newtheorem{prop}{Proposition}{\bf}{\em}
\newtheorem{lem}{Lemma}{\bf}{\em}
\newtheorem{rem}{Remark}{\bf}{\em}
{\bf}{\em}

\def\bpm{\begin{pmatrix}}
\def\epm{\end{pmatrix}}

\def\bel{\begin{lem}}
\def\eel{\end{lem}}
\def\ber{\begin{rem}}
\def\eer{\end{rem}}
\def\bep{\begin{prop}}
\def\eep{\end{prop}}

\def\<{\langle}
\def\>{\rangle}

\def\eg/{e.g.}
\def\ie/{i.e.}

\makeatletter
\def\Wtilde#1{\mathop{\vbox{\m@th\ialign{##\crcr\noalign{\kern3\p@}%
      \sortoftildefill\crcr\noalign{\kern3\p@\nointerlineskip}%
      $\hfil\displaystyle{#1}\hfil$\crcr}}}\limits}

\def\sortoftildefill{$\m@th \setbox\z@\hbox{$\braceld$}%
  \braceld\leaders\vrule \@height\ht\z@ \@depth\z@\hfill\braceru$}

\allowdisplaybreaks[3]
\begin{document}
\renewcommand{\baselinestretch}{1.5}\normalsize
\title[Higher-Order Nonlinear Schr{\"o}dinger Equation]{\emph{}\\ Quasi-Gramian Solution of a Noncommutative Extension of the Higher-Order Nonlinear Schr{\"o}dinger Equation }

\author[H. W. A. Riaz
\lowercase{and}
J. Lin
]{\emph{}\\
H. W. A. Riaz$^{1}$
\lowercase{and}
J. Lin$^1$
\\
\lowercase{\scshape{
${}^1$
Department of Physics, Zhejiang Normal University, Jinhua 321004,
PR China
}}
    }

\thanks{ wajahat@zjnu.edu.cn,  linji@zjnu.edu.cn}

\maketitle

\begin{abstract}
The nonlinear Schr{\"o}dinger (NLS) equation, which incorporates higher-order dispersive terms, is widely employed in the theoretical analysis of various physical phenomena. In this study, we explore the non-commutative extension of the higher-order NLS equation (HNLS). We treat real or complex-valued functions, such as $g_1=g_1 (x,\;t)$ and $g_2= g_2 (x ,\;t)$, as non-commutative, and employ the Lax pair associated with the evolution equation as in the commutation case. We derive the quasi-Gramian solution of the system by employing a binary Darboux transformation (DT). Moreover, the solution can be used to study the stability of plane waves and to understand the generation of periodic patterns in the context of modulational instability.
\end{abstract}
\section{Introduction}
The nonlinear Schr{\"o}dinger (NLS) equation, which incorporates higher-order dispersive terms, is widely used in the theoretical analysis of various physical phenomena, including nonlinear optics, molecular systems, and fluid dynamics \cite{ablowitz2004discrete,kivshar2003optical,agrawal2000nonlinear}. With the addition of fourth-order terms, known as the Lakshmanan-Porsezian-Daniel (LPD) equation, it describes higher-order molecular excitations with quadruple-quadruple coefficients and possesses integrability \cite{daniel1995davydov,lakshmanan1988effect,porsezian1992integrability}. Lakshmanan et al. investigated its application to study nonlinear spin excitations involving bilinear and biquadratic interactions \cite{lakshmanan1988effect}. In recent years, Ankiewicz et al. introduced a further extension of the NLSE by incorporating third-order (odd) and fourth-order (even) dispersion terms \cite{ankiewicz2014extended}. The integrability of this extended NLSE, with certain parameter values, was confirmed in Ref. \cite{ankiewicz2014higher}, where Lax operators were introduced. We now write this equation appears in the mentioned references with some modification as
\begin{eqnarray}\label{mainNLS}
 &&\mathrm{i} u_{t}+\alpha_{2} \left(u_{xx}+2 u |u|^{2} \right)+\mathrm{i} \alpha_{1} \left(u_{xxx}+6 u_{x} |u|^{2}\right)+\gamma  (u_{xxxx}+6 \bar{u} u_{x}^{2} +4 u |u_{x}|^{2} \\ \notag && +8 |u|^{2} u_{xx}+2 u^{2} \bar{u}_{xx}+6 u |u|^{4} )=0.  
\end{eqnarray}
where $u = u(x, \;t)$ is a complex-valued scalar function, and $\bar{u}$ represents its complex conjugate. This equation includes several particular cases, such as the standard nonlinear Schrödinger equation (NLSE) with $\alpha_{1} = \gamma = 0$ \cite{lakshmanan1977continuum}, the Hirota equation with $\gamma=0$ \cite{hirota1973exact}, and the Lakshmanan-Porsezian-Daniel equation with $\alpha_{1}=0$ \cite{lakshmanan1988effect}.

In this study, we explore the non-commutative extension of the higher-order NLS (HNLS) equation (\ref{mainNLS}). Noncommutative integrable systems have received considerable attention for their relevance in quantum field theories, $D$-brane dynamics, and string theories \cite{minwalla2000noncommutative,furuta2000ultraviolet,Seiberg1999}. Non-commutativity often arises from phase-space quantization, introducing non-commutativity among independent variables through a star product \cite{dimakis2000bicomplexes,lechtenfeld2001noncommutative}. Our approach to inducing non-commutativity in a given nonlinear evolution equation parallels the methods employed by Lechtenfeld \textit{et al.} \cite{Lechtenfeld2005}, Gilson and Nimmo \cite{Gilson2007}, and Gilson and Macfarlane \cite{Gilson2009} for the non-commutative generalization of the sine-Gordon, Kadomtsev-Petviashvili (KP), and Davey-Stewartson (DS) equations, respectively.

We adopt a systematic method to extend the chosen equation to its non-commutative form, without explicitly specifying the nature of non-commutativity. We consider real or complex-valued functions, such as $g_{1}=g_{1}(x,\;t)$ and $g_{2}=g_{2}(x,\;t)$, as non-commutative and take advantage of the same Lax pair as in the commutative scenario to describe the equation of non-linear evolution.

In this paper, we investigate a non-commutative (nc) version of the HNLS equation.
We define the Lax pair for the nc-HNLS equation in this context. To find solutions to the nc-HNLS equation, we construct the Darboux matrix and the binary Darboux matrix. We present explicit quasi-Gramian solutions for the non-commutative fields of the nc-HNLS equation, which, after reducing the non-commutativity limit, can be reduced to a ratio of Gramian solutions.

\section{Modulation instability}
For analyzing the modulation instability, we give a plane-wave solution of the system (\ref{mainNLS})
\begin{equation}
u(x, t) = ce^{i(6c^4 \gamma + 2\alpha_2 c^2) t }, \label{planewave}
\end{equation}
The solution provided by equation (\ref{planewave}) holds significant importance in the realm of optics, particularly within the context of the nonlinear Schr{\"o}dinger equation. This solution represents a wave with constant amplitude that undergoes a non-linear evolution over time. The dynamics are determined by parameters such as the amplitude $c$, the constant $\gamma$, and $\alpha_2$. This solution's application extends to studying the stability of plane waves and comprehending the generation of periodic patterns through modulational instability. It serves as a prime example of how the nonlinear Schr{\"o}dinger equation can lead to complex behavior in optical systems, thereby making it a crucial area of research in this field.

An approach to assess the stability of the plane wave solution involves introducing perturbations to the solution and examining the linearized evolution of these perturbations. To simplify the analysis, the common phase can be factored out of the equation. This leads to a first-order ordinary differential equation (ODE) that couples the complex field with its complex conjugate as a result of the perturbation. Substituting the perturbed function $v(x,\;t)$ into equation (\ref{planewave}), we obtain
    \begin{equation}
    u(x, t) = (c+v(x,\;t))e^{i(6c^4 \gamma + 2\alpha_2 c^2) t }, \label{planewavepert}
\end{equation}
Substituting (\ref{planewavepert}) into (\ref{mainNLS}) and after linearization, we have
\begin{equation}
    i v_t + \alpha_2 v_{xx} + i \alpha_1 (v_{xxx} + 6 c^2 v_{x}) + \gamma v_{xxxx} + 2 \gamma c^2 (4 v_{xx} + \bar{v}_{xx}) + c^2 (12 \gamma c^2 + 2 \alpha_2) (v + \bar{v}) = 0, \label{pert1} 
\end{equation}

In order to analyze the stability of the plane wave solution, the Fourier transform of the equation is taken. This results in a first-order ODE that governs the real and imaginary parts of evolution. The stability of the solution can then be determined by looking at the eigenvalues of this ODE. In particular, the eigenvalues represent the exponent in the time evolution of the solution. Thus, the Fourier transform of the evolution equation (\ref{pert1}) is
\begin{equation}
    i \hat{v}_t - \frac{\alpha_2}{2} k^2 \hat{v} -  \alpha_1 k(-k^2 + 6 c^2 )\hat{v} + \gamma k^4 \hat{v} - 2 \gamma c^2 k^2 (4 \hat{v} + \bar{\hat{v}}) + c^2 (12 \gamma c^2 + 2 \alpha_2) (\hat{v} + \bar{\hat{v}}) = 0, \label{pert2} 
\end{equation}

\begin{figure}[H]
         \centering
         \includegraphics[width=0.4\textwidth]{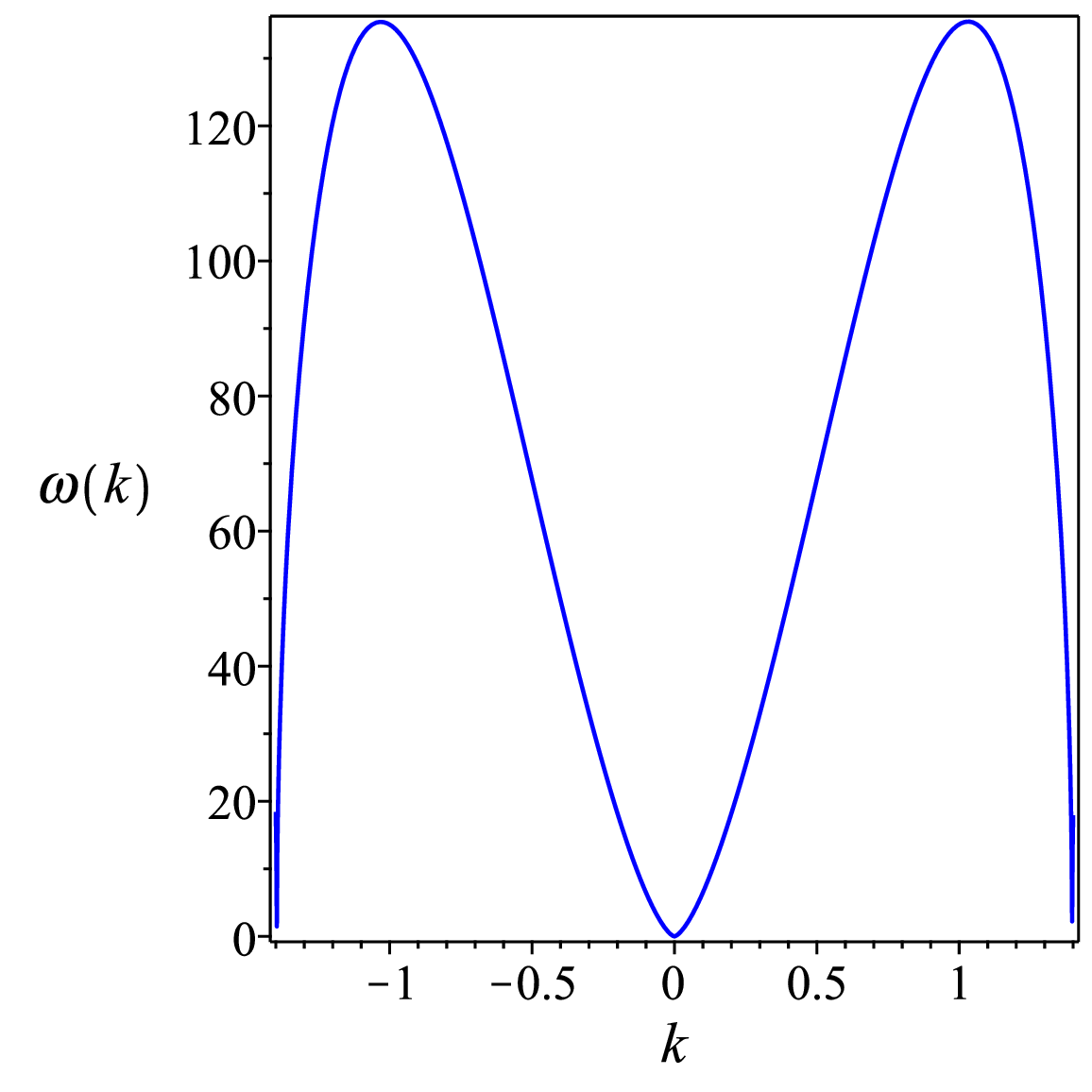}
         \caption{}
         \label{fig:figMI}
     \end{figure}

The linear evolution equation for $\hat{v}$ can be evaluated by separating it into its real and imaginary components. Thus, for $\hat{v} = v_1 + i v_2$, we have system of differential equation
\begin{equation}
    y_t = \left[\begin{array}{cc}
0  & \beta-6 \gamma  \,c^{2} k^{2} 
\\
 -\beta -10 \gamma  \,c^{2} k^{2}+2 c^{2} \left(12 c^{2} \gamma +2 \alpha_2 \right) & 0  
\end{array}\right] y ,
\end{equation}
where $\beta = \frac{\alpha2 \,k^{2}}{2}+\alpha_1 k \left(6 c^{2}-k^{2}\right)-\gamma  \,k^{4}$, and $ y = [\begin{array}{cc} v_1 & v_2\end{array} ]^{T}$. One can evaluate the stability of a system by analyzing exponential solutions in the form of $y = \nu e^{ \omega t }$, which leads to an eigenvalue problem where the eigenvalues are then given by
\begin{eqnarray}\label{MI}
    &&\omega(k) = {\frac{| k |}{2} \sqrt{\left(\alpha_2  \left(-8 c^{2}+k^{2}\right)-2 \alpha_1 k \left(-6 c^{2}+k^{2}\right)-2 \gamma  \left(24 c^{4}-10 c^{2} k^{2}+k^{4}\right)\right)\beta_1}}, \\
    && \beta_1 = k \alpha_2 -2 \alpha_1 \left(-6 c^{2}+k^{2}\right)-2 \gamma  k \left(6 c^{2}+k^{2}\right). \notag
\end{eqnarray}
The plot of the equation (\ref{MI}) is shown in Fig. \ref{fig:figMI}
The stability of the solution becomes evident when examining a graphical representation of the real parts of eigenvalues plotted against different frequencies. If the real part is positive, the solution will exhibit growth; conversely, negative values indicate decay. The overall stability is established by observing whether the real eigenvalue remains positive or negative across various frequency ranges. This phenomenon is referred to as modulational instability. 

\section{Non-commutative HNLS equation}
The spectral problem associated with the nc-HNLS equation is given by 
\begin{eqnarray}
    && \Gamma = \partial_x - \lambda \mathcal{J} - \mathcal{U}, \label{lax1} \\
    && \Delta = \partial_t - \mathcal{B} - \mathcal{V}_p, \label{lax2}
\end{eqnarray}
where
 \begin{eqnarray}\label{matrices}
 && \mathcal{J} = \left[\begin{array}{cc}
\mathrm{-I} & 0 
\\
 0 & \mathrm{I} 
\end{array}\right], \quad \mathcal{U} = \left[\begin{array}{cc}
0 & {u} 
\\
 -{\mathit{u^{\dag}}} & 0 
\end{array}\right],   \quad \mathcal{V}_p = \gamma \left[\begin{array}{cc}
\mathcal{V}_1 & \mathcal{V}_2 
\\
 \mathcal{V}_3 & \mathcal{V}_4 
\end{array}\right],
\end{eqnarray}
\begin{eqnarray}\label{Bmatrix}
&&  \mathcal{B} = \small{\begin{pmatrix}
 \rho_1 + \rho_2 {u} {\mathit{u^{\dag}}} -
\alpha_{1} \left(\left(\frac{\partial}{\partial x} {u}\right) {\mathit{u^{\dag}}}- {u} \left(\frac{\partial}{\partial x} {\mathit{u^{\dag}}}\right)\right) & \mathcal{A}_2 
\\
 \mathcal{A}_1 & -\rho_1 -\rho_2 {\mathit{u^{\dag}}}{u}-
\alpha_{1} \left(-{\mathit{u^{\dag}}} \left(\frac{\partial}{\partial x} {u}\right)+\left(\frac{\partial}{\partial x} {\mathit{u^{\dag}}}\right) {u}\right) 
\end{pmatrix}}, \\
&& \mathcal{A}_1 = \small{-4 \lambda^{2} \alpha_{1} {\mathit{u^{\dag}}}+2 \lambda  \left(-\alpha_{2} {\mathit{u^{\dag}}}+\mathrm{I} \alpha_{1} \left(\frac{\partial}{\partial x}{\mathit{u^{\dag}}}\right)\right)+\mathrm{I} \alpha_{2} \left(\frac{\partial}{\partial x}{\mathit{u^{\dag}}}\right)-
\alpha_{1} \left(-2 {\mathit{u^{\dag}}} {u} {\mathit{u^{\dag}}}-\frac{\partial^{2}}{\partial x^{2}} {\mathit{u^{\dag}}}\right)}, \notag \\
&& \mathcal{A}_2 = 4 \lambda^{2} \alpha_{1}  {u} + 2 \lambda  \left(\alpha_{2}  {u}+\mathrm{I} \alpha_{1} \left(\frac{\partial}{\partial x} {u}\right)\right)+\mathrm{I} \alpha_{2} \left(\frac{\partial}{\partial x} {u}\right)-\alpha_{1} \left(2 {u} {\mathit{u^{\dag}}} {u}+\frac{\partial^{2}}{\partial x^{2}}{u}\right), \notag
\end{eqnarray}
\begin{eqnarray}
  && \mathcal{V}_1 =
  \mathrm{I} \left(\left(\frac{\partial^{2}}{\partial x^{2}}{u}\right) {\mathit{u^{\dag}}} + {u} \left(\frac{\partial^{2}}{\partial x^{2}}{\mathit{u^{\dag}}}\right)-\left(\frac{\partial}{\partial x}{u}\right) \left(\frac{\partial}{\partial x}{\mathit{u^{\dag}}}\right)\right)- 
2 \lambda  \left({u} \left(\frac{\partial}{\partial x}{\mathit{u^{\dag}}}\right)-\left(\frac{\partial}{\partial x} {u}\right) {\mathit{u^{\dag}}}\right), \notag \\
&& \mathcal{V}_2 = -4 \,\mathrm{I} \lambda^{2} \left(\frac{\partial}{\partial x}{u}\right)+3 \,\mathrm{I} \left({u} {\mathit{u^{\dag}}} \left(\frac{\partial}{\partial x}{u}\right)+\left(\frac{\partial}{\partial x}{u}\right) {\mathit{u^{\dag}}} {u}\right)+\mathrm{I} \left(\frac{\partial^{3}}{\partial x^{3}}{u}\right)
+2 \lambda  \left(\frac{\partial^{2}}{\partial x^{2}}{u}\right), \notag
\\
&& \mathcal{V}_3 = -4 \,\mathrm{I} \lambda^{2} \left(\frac{\partial}{\partial x} {\mathit{u^{\dag}}}\right)+3 \,\mathrm{I} \left({\mathit{u^{\dag}}} {u} \left(\frac{\partial}{\partial x}{\mathit{u^{\dag}}}\right)+\left(\frac{\partial}{\partial x}{\mathit{u^{\dag}}}\right) {u} {\mathit{u^{\dag}}}\right)+
\mathrm{I} \left(\frac{\partial^{3}}{\partial x^{3}}{\mathit{u^{\dag}}}\right)-2 \lambda  \left(\frac{\partial^{2}}{\partial x^{2}}{\mathit{u^{\dag}}}\right), \notag \\ 
&& \mathcal{V}_4 = -\mathrm{I} \left(\left(\frac{\partial^{2}}{\partial x^{2}} {\mathit{u^{\dag}}}\right) {u}+{\mathit{u^{\dag}}} \left(\frac{\partial^{2}}{\partial x^{2}} {u}\right)-\left(\frac{\partial}{\partial x}{\mathit{u^{\dag}}}\right) \left(\frac{\partial}{\partial x}{u}\right)\right)+
2 \lambda  \left(-{\mathit{u^{\dag}}} \left(\frac{\partial}{\partial x} {u}\right)+\left(\frac{\partial}{\partial x} {\mathit{u^{\dag}}}\right) {u}\right), \notag
\end{eqnarray}
where $\rho_1 = -4 \,\mathrm{I} \alpha_{1} \lambda^{3}-2 \,\mathrm{I} \lambda^{2} \alpha_{2},\; \rho_2 = \,\mathrm{I} (2 \lambda  \alpha_{1} + \alpha_{2} )$, $\rho_3 = 3 \,\mathrm{I} {u} {\mathit{u^{\dag}}}{u}{\mathit{u^{\dag}}} -4 \,\mathrm{I} \lambda^{2} {u} {\mathit{u^{\dag}}}+8 \,\mathrm{I} \lambda^{4} ,\; \rho_4 = -3 \,\mathrm{I} {\mathit{u^{\dag}}} {u}  {\mathit{u^{\dag}}} {u} +4 \,\mathrm{I} \lambda^{2} {\mathit{u^{\dag}}} {u}-8 \,\mathrm{I} \lambda^{4} $, $\rho_5 = -8 \lambda^{3} {u} +4 \lambda  {u} {\mathit{u^{\dag}}}{u}, \; \rho_6 = 8 \lambda^{3} {\mathit{u^{\dag}}}  -4 \lambda  {\mathit{u^{\dag}}} {u} {\mathit{u^{\dag}}} $. The equation of motion for the system can be derived by setting the commutator of $\Gamma$ and $\Delta$ equal to zero and equating the coefficients at $\lambda$
\begin{eqnarray}\label{EOM}
    &&i {u}_{t} + i \left(  {u}_{xxx} + 3 ( {u}_{x} {\mathit{u^\dag}} {u}+ {u} {\mathit{u^\dag}}  {u}_{x}) \right) \alpha_{1} +\left(2 {u} {\mathit{u^\dag}} {u}+ {u}_{xx} \right) \alpha_{2} 
    +({u}_{xxxx}  \\
    && + 2  ({u}_{x}  {\mathit{u^\dag}}_{x}  {u} + {u} {\mathit{u^\dag}}_{x}  {u}_{x} + {u}  {\mathit{u^\dag}}_{xx}  {u})  +  4  ({u}_{xx}  {\mathit{u^\dag}}  {u} + {u} {\mathit{u^\dag}} {u}_{xx}) + 6 ( {u}_{x} {\mathit{u^\dag}}  {u}_{x} +   {u} {\mathit{u^\dag}} {u} {\mathit{u^\dag}} {u} )  ) \gamma  =0, \notag
\end{eqnarray}
where $u = u(x, \; t)$ is an nc object, $^{\dag}$ denotes the adjoint (Hermitian conjugate), $\alpha_1, \; \alpha_2$ and $\gamma$ are real parameters, and $\lambda$ is a spectral parameter (real or complex). The equation presented in (\ref{EOM}) is a non-commutative generalization of the HNLS equation, as given in (\ref{mainNLS}). This equation exhibits several interesting properties. For instance, when both $\alpha_2$ and $\gamma$ are zero, it reduces to an nc generalization of the complex modified Korteweg–de Vries (KdV) equation and to the standard modified KdV equation for real-valued $u$. Moreover, setting $\gamma$ to zero results in an nc generalization of the Hirota equation, while setting $\alpha_1$ and $\alpha_2$ to zero simultaneously yields the well-known Lakshmanan–Porsezian–Daniel (LPD) equation. Finally, when $\alpha_1$ and $\gamma$ are set to zero, the equation reduces to an nc generalization of the nonlinear Schrödinger (NLS) equation. After relaxing the noncommutativity condition, Eq. (\ref{EOM}) corresponds to the commutative counterpart. The spectral problem linked with (\ref{mainNLS}) remains the same as that of (\ref{lax1}) and (\ref{lax2}), with the exception that $u$ and $\bar{u}$ are now perceived as commutative functions.

\subsection{Quasideterminants}
In non-commutative algebra, quasideterminants serve as a replacement for ordinary determinants of matrices. They hold a similar significance in non-commutative algebra as ordinary determinants do in commutative algebra and have found vast applications in the domain of non-commutative integrable systems \cite{Gilson2007,etingof97,riaz2018}.

The quasideterminant $|M|_{ij}$ for $i,\;j = 1,\; ...,\;n$ of an $n\times n$ matrix over a non-commutative ring R, expanded about the matrix $m_{ij}$, is defined as
\begin{equation}
\left|M\right|_{ij} = \left\vert
\begin{array}{cc}
M^{ij} & c^{i}_{j} \\
r^{j}_{i} & \fbox{$m_{ij}$}
\end{array}%
\right\vert = m_{ij} - r^{j}_{i} \left(M^{ij}\right)^{-1}c^{i}_{j},
\label{def}
\end{equation}
where $m_{ij}$ is referred to as the expansion point and represents the $ij$-th entry of $M$, $r^{j}_{i}$ denotes the $i$-th row of $M$ without the $j$-th entry, $c^{i}_{j}$ represents the $j$-th column of $M$ without the $i$-th row, and $M^{ij}$ is the submatrix of $M$ obtained by removing the $i$-th row and the $j$-th column from $M$.

Quasideterminants are not merely a generalization of usual commutative determinants but are also related to inverse matrices. The inverse
of a matrix $M = \left(
                   \begin{smallmatrix}
                     m_{11} & m_{12} \\
                     m_{21} & m_{22}
                   \end{smallmatrix}
                 \right)
$ is defined as
\begin{eqnarray}
M^{-1} = \left(|M|_{ji}^{-1}\right) = \left(
      \begin{array}{cc}
        \left\vert\begin{array}{cc}
         \fbox{$m_{11}$} & m_{12} \\
          m_{21} & m_{22}
          \end{array}%
           \right\vert^{-1}  & \left\vert\begin{array}{cc}
         {m_{11}} & \fbox{$m_{12}$} \\
          {m_{21}} & m_{22}
          \end{array}%
           \right\vert^{-1} \\
        \left\vert\begin{array}{cc}
         {m_{11}} & m_{12} \\
          \fbox{$m_{21}$} & m_{22}
          \end{array}%
           \right\vert^{-1} & \left\vert\begin{array}{cc}
         {m_{11}} & m_{12} \\
          m_{21} & \fbox{$m_{22}$}
          \end{array}%
           \right\vert^{-1}
      \end{array}
    \right).
\end{eqnarray}

\section{Darboux transformation}
In this section, a Darboux transformation is introduced for the system of nc-HNLS equation (\ref{EOM}) through the definition of the Darboux matrix
\begin{eqnarray}
D (\mathcal{Y}) = \lambda I - \mathcal{Y} \Lambda \mathcal{Y}^{-1}, \label{D1}    
\end{eqnarray}
and the Lax operators $\Gamma$ and $\Delta$ 
\begin{equation}
    \Gamma = \partial_x - \lambda \mathcal{J} - \mathcal{U},  \quad \Delta = \partial_t - \mathcal{B} - \mathcal{V}_p. \label{RQ}
\end{equation}
The spectral parameter $\lambda $, which can be real or complex, is incorporated along with the constant $q \times q$ matrix $\Lambda$, and the nc objects $\mathcal{U},\; \mathcal{V}_p$, and $\mathcal{B}$ from (\ref{matrices}) and (\ref{Bmatrix}) respectively, are utilized as entries in the Lax operators. Now, consider a function $\varphi = \varphi(x,\;t)$ that is an eigenfunction of the Lax operators $\Gamma$ and $\Delta$ such that $\Gamma(\varphi) = 0$ and $\Delta(\varphi) = 0$. We can define a new function $\Tilde{\varphi}$ using the Darboux matrix $D(\mathcal{Y})$:
\begin{eqnarray}
\Tilde \varphi &:=& D_{\mathcal{Y}}(\varphi) , \notag \\
&=& \lambda \varphi - \mathcal{Y} \Lambda \mathcal{Y}^{-1} \phi = \left\vert
\begin{array}{cc}
  \mathcal{Y} & \varphi \\
  \mathcal{Y} \Lambda & \fbox{$\lambda \varphi$}
\end{array}%
\right\vert \label{Dquasi1}
\end{eqnarray}
The function $\Tilde{\varphi}$ is a generic function of the new operators $\Tilde{ \Gamma}
= D_{\mathcal{Y}}  \Gamma D_{\mathcal{Y}}^{-1}$ and $\Tilde{ \Delta} =
D_{\mathcal{Y}} \Delta D_{\mathcal{Y}}^{-1}$.
Next, we define $\mathcal{Y}_{[1]} = \mathcal{Y}_1$ and $\varphi_{[1]} = \varphi$ as generic eigenfunctions of $\Gamma_{[1]} = \Gamma$ and $\Delta_{[1]} = \Delta$, where $\mathcal{Y}_1$ and $\varphi_1$ are $2 \times 2$ matrices. We then define $\varphi_{[2]}=D_{\mathcal{Y}_{[1]}}\left(\varphi_{[1]}\right)$ and $\mathcal{Y}_{[2]}=\varphi_{[2]}|_{\varphi \rightarrow \mathcal{Y}_{2}}$ to be eigenfunctions of the new operators ${ \Gamma}_{[2]} =
D_{\mathcal{Y}_{[1]}}{ \Gamma}_{[1]} D_{\mathcal{Y}_{[1]}}^{-1}$ and ${
\Delta}_{[2]} = D_{\mathcal{Y}_{[1]}}{ \Delta}_{[1]} D_{\mathcal{Y}_{[1]}}^{-1}$. 

We define $\mathcal{Y}_{i},\;i=1,\;...,\;n$ (where each $\mathcal{Y}_{i}$ is a
matrix of size $2\times2$) be the set of particular eigenfunctions
of $ \Gamma$ and $ \Delta$. In the non-commutative case, we consider each entry of $\mathcal{Y}_{i}$ as a matrix. We define the $(n+1)$th eigenfunction as  $\varphi_{[n+1]}=D_{\mathcal{Y}_{[n]}}\left(\varphi_{[n]}\right)$, which is a generic eigenfunction of the new operators ${ \Gamma}_{[n+1]} = D_{\mathcal{Y}_{[n]}}{ \Gamma}_{[n]}
D_{\mathcal{Y}_{[n]}}^{-1}$, and $
{ \Delta}_{[n+1]} = D_{\mathcal{Y}_{[n]}}{ \Delta}_{[n]}
D_{\mathcal{Y}_{[n]}}^{-1}$. Here, $\mathcal{Y}_{[n]}$ is a DT from $\Gamma_{[n]}$ to $\Gamma_{[n+1]}$ and $\Delta_{[n]}$ to $\Delta_{[n+1]}$. Thus, the $n$th-order DT is given by $\varphi_{[n+1]}=D_{\mathcal{Y}_{[n]}}\left(\varphi_{[n]}\right) = \lambda \varphi_{[n]} - \mathcal{Y}_{[n]} \Lambda_{n} \mathcal{Y}_{[n]}^{-1}
\varphi_{[n]}$, where $\mathcal{Y}_{[j]}=\varphi_{[j]}|_{\varphi \rightarrow \mathcal{Y}_{j}}$. Let us define a matrix $\Xi = (\mathcal{Y}_{1},\;...,\; \mathcal{Y}_{n})$ comprising of the eigenfunctions $\mathcal{Y}_i ,\; i = 1, \; . . . ,\; n$, and set $\varphi_{[1]} = \varphi$. Then, we can represent the $n$th iteration of the DT in quasideterminant form as follows:
\begin{equation}
\varphi_{[n+1]} = \left\vert
\begin{array}{cc}
  \Xi & \varphi \\
  \vdots & \vdots \\
  \Xi^{(n-1)} & \varphi^{(n-1)} \\
  \Xi^{(n)} & \fbox{$\varphi^{(n)}$}
\end{array}%
\right\vert, \label{phin11}
\end{equation}
Here, $\varphi^{(n)}=\lambda^{n}\varphi$ and $\Xi^{(n)}=\Xi
{\Lambda}^{n}$., where each $\Lambda^{i},\;i=1,\;...,\;n$, is a constant matrix. Hence, we expressed a quasideterminant formula for $\varphi_{n+1}$ in terms of the known eigenfunctions $\mathcal{Y}_{i},\;i=1,\;...,\;n$ and the eigenfunction $\varphi$ of the “seed” Lax pair $\Gamma = \Gamma_1$, $\Delta = \Delta_1$.

\section{Quasi-Wronskian solutions}
In the upcoming analysis, we will examine how the DT $D_{\mathcal{Y}} = \lambda I -
\mathcal{Y} \Lambda \mathcal{Y}^{-1}$ affects the Lax operator $\Gamma = \Gamma_1$, where $\mathcal{Y}$ is an eigenfunction of $\Gamma$ (since $\Gamma(\mathcal{Y}) = 0$ by definition) and $\Lambda$ is an eigenvalue matrix. It is important to note that the same results apply to the operator $\Delta = \Delta_1$. As a result of this transformation, the operator $\Gamma$ is converted to a new operator $\Tilde{\Gamma} = \Gamma_{[2]}$, which can be expressed as $\Tilde{ \Gamma} = D_{\mathcal{Y}}{ \Gamma}D_{\mathcal{Y}}^{-1}$. By substituting (\ref{lax1}) and (\ref{D1}) into the latter equation and equating the coefficients at $\lambda^j$, we obtain two equations,
\begin{eqnarray}
\mathcal{U}_{[2]} - \mathcal{U} - [\mathcal{J},\; \mathcal{Y} \Lambda \mathcal{Y}^{-1}]  =0, \label{R1}
\end{eqnarray}
and
\begin{equation}
-\partial_{x} (\mathcal{Y} \Lambda \mathcal{Y}^{-1}) + [\mathcal{U},\; \mathcal{Y} \Lambda \mathcal{Y}^{-1}] +
[\mathcal{J},\; \mathcal{Y} \Lambda \mathcal{Y}^{-1}] \mathcal{Y} \Lambda \mathcal{Y}^{-1}=0. \label{cond1}
\end{equation}
To confirm the validity of equation (\ref{cond1}), we express equation (\ref{lax1}) using a particular eigenfunction $\mathcal{Y}$ as $\mathcal{Y}_{x} =  \mathcal{J} \mathcal{Y} \Lambda + \mathcal{U} \mathcal{Y}$. By utilizing this equation, we can easily check that the condition expressed in (\ref{cond1}) is satisfied.
To simplify the notation, a matrix $\mathcal{F}$ is introduced such that $\mathcal{U} = [\mathcal{F},\;\mathcal{J}]$. This equation is satisfied if $\mathcal{F} = \frac{1}{2\dot\imath}
                           \left(\begin{smallmatrix}
                             0 & {u} \\
                             {u}^{\dag} & 0
                           \end{smallmatrix}
                         \right)$. Then, equation (\ref{R1}) with $\mathcal{U} = [\mathcal{F},\;\mathcal{J}]$ can be used to obtain $\mathcal{F}_{[2]} = \mathcal{F} - \mathcal{Y}^{(1)}\mathcal{Y}^{-1}$, where $\mathcal{Y}^{(1)}$ is defined as $\mathcal{Y} \Lambda$. After $n$ repeated applications of the DT $D_\mathcal{Y}$, we have
\begin{eqnarray}
\mathcal{F}_{[n+1]} = \mathcal{F}_{[n]} - \mathcal{Y}^{(1)}_{[n]}\mathcal{Y}^{-1}_{[n]}= \mathcal{F} - \sum_{k=1}^{n}\mathcal{Y}^{(1)}_{[k]}\mathcal{Y}^{-1}_{[k]}, \label{Un}
\end{eqnarray}
where $\mathcal{F}_{[1]}=\mathcal{F},\; \mathcal{Y}_{[1]}=\mathcal{Y}_{1}=\mathcal{Y}$ and
$\Lambda_{1}=\Lambda$.

Because our nc-HNLS equation (\ref{EOM}) is expressed in terms of $u$ and $u^{\dag}$, it is more appropriate to express
the quasi-Wronskian solution in terms of these objects. For this, we express each $\mathcal{Y}_{i},\;i=1,\;...,\;n$ as a $2
\times 2$ matrix as $\mathcal{Y}_{i} = \left(
               \begin{smallmatrix}
                 \varphi_{2i-1} & \varphi_{2i} \\
                 \chi_{2i-1} & \chi_{2i}  
               \end{smallmatrix}
             \right)$. For $\varphi=\varphi(x,\;t)$ and $\chi=\chi(x,\;t)$, we can express
$\mathcal{F}_{[n+1]}$ as
\begin{equation}
\mathcal{F}_{[n+1]} = \mathcal{F} + \left(\begin{array}{cc}
                                    \left\vert
\begin{array}{cc}
\widehat \Xi & f_{2n-1} \\
\varphi^{(n)} & \fbox{$ 0 $}%
\end{array}%
\right\vert & \left\vert
\begin{array}{cc}
\widehat \Xi & f_{2n} \\
\varphi^{(n)} & \fbox{$ 0 $}%
\end{array}%
\right\vert
\\ \\
\left\vert
\begin{array}{cc}
\widehat \Xi & f_{2n-1} \\
\chi^{(n)} & \fbox{$ 0 $}%
\end{array}%
\right\vert & \left\vert
\begin{array}{cc}
\widehat \Xi & f_{2n} \\
\chi^{(n)} & \fbox{$ 0 $}%
\end{array}%
\right\vert
\end{array}
\right), \label{Fn1}
\end{equation}
where $\widehat \Xi = (\mathcal{Y}_{j}^{(i-1)})_{i,\;j=1,\;...,\;n}$ is the
$2n \times 2n$ matrix of $\mathcal{Y}_{1},\;...,\;\mathcal{Y}_{n}$, and $f_{2n-1}$
and $f_{2n}$ are the column vectors $2n \times 1$ with a 1 in the
$(2n-1)$th and $(2n)$th row, respectively, and zeros elsewhere,
while $\varphi^{(n)} = \left( \varphi_{1}^{(n)},\;...,\;
\varphi_{2n}^{(n)} \right) ,\; \chi^{(n)} = \left(
\chi_{1}^{(n)},\;...,\; \chi_{2n}^{(n)} \right)$ denote the $1
\times 2n$ row vectors. We can express the quasi-Wronskian solutions for $u$ and $u^{\dag}$ by utilizing $\mathcal{F}=\frac{1}{2\dot\imath}\left(
                                      \begin{smallmatrix}
                                        0 & u \\
                                        u^{\dagger} & 0 \\
                                      \end{smallmatrix}
                                    \right)$ which leads to:
\begin{equation}
u_{[n+1]} = u + 2\dot\imath \left\vert
\begin{array}{cc}
\widehat\Xi & f_{2n} \\
\varphi^{(n)} & \fbox{$ 0 $}%
\end{array}%
\right\vert, \qquad  u_{[n+1]}^{\dagger} = u^{\dagger} + 2\dot\imath
\left\vert
\begin{array}{cc}
\widehat\Xi & f_{2n-1} \\
\chi^{(n)} & \fbox{$ 0 $}%
\end{array}%
\right\vert.
\end{equation}
We proceed in the next section to construct the binary DT for the nc-HNLS equation, using a strategy similar to that employed in \cite{nimmo2000applications}.

\section{Binary Darboux transformation}
We introduce $\mathcal{Y}_{1},\;...,\; \mathcal{Y}_{n}$ as eigenfunctions of the Lax operators $\Gamma$ and $\Delta$, and $\mathcal{Z}_{1},\;...,\; \mathcal{Z}_{n}$ as 
eigenfunctions of the adjoint Lax operators $\Gamma^{\dag}$ and $\Delta^\dag$. Assuming $\varphi_{[1]} = \varphi$ to be a generic 
eigenfunction of the Lax operators $\Gamma$ and $\Delta$, and $\psi_{[1]} = \psi$ to be a generic eigenfunction of the adjoint Lax 
operators $\Gamma^{\dag}$ and $\Delta^{\dag}$, we define the binary DT and its adjoint as
\begin{equation}
D_{\mathcal{Y}_{[1]},\;\mathcal{Z}_{[1]}} = I -
\mathcal{Y}_{[1]}\Upsilon(\mathcal{Y}_{[1]},\;\mathcal{Z}_{[1]})^{-1} \Omega^{-\dag}
\mathcal{Z}_{[1]}^{\dag}, \label{BDT1}
\end{equation}
\begin{equation}
D_{\mathcal{Y}_{[1]},\;\mathcal{Z}_{[1]}}^{\dag} = I -
\mathcal{Z}_{[1]}\Upsilon(\mathcal{Y}_{[1]},\;\mathcal{Z}_{[1]})^{-\dag} \Lambda^{-\dag}
\mathcal{Y}_{[1]}^{\dag}. \label{BDT2}
\end{equation}
In the context of the binary DT, we use $\mathcal{Y}_{[1]} = \mathcal{Y}_1$ as the initial eigenfunction that characterizes the transformation from the Lax operators $\Gamma$ and $\Delta$ to the new operators $\Tilde{\Gamma}$ and $\Tilde{\Delta}$. Similarly, we define $\mathcal{Z}_{[1]} = \mathcal{Z}_1$ to represent the adjoint transformation, where the potential $\Upsilon$ 
\begin{eqnarray}
&&\Omega^{\dag} \Upsilon (\mathcal{Y}_{[1]},\;\mathcal{Z}_{[1]}) + \Upsilon
(\mathcal{Y}_{[1]},\;\mathcal{Z}_{[1]}) \Lambda = \mathcal{Z}_{[1]}^{\dag}\mathcal{Y}_{[1]},
\label{omega1} \\
&& \left(\lambda I + \Omega^{\dag}\right) \Upsilon
(\varphi_{[1]},\;\mathcal{Z}_{[1]}) = \mathcal{Z}_{[1]}^{\dag}\varphi_{[1]},
\label{omega2} \\
&& \Upsilon (\mathcal{Y}_{[1]},\;\psi_{[1]})\left(\mu^{\dag} I + \Lambda
\right) = \psi_{[1]}^{\dag} \mathcal{Y}_{[1]}. \label{omega3}
\end{eqnarray}
The transformed operators $\widetilde{ \Gamma} = 
\Gamma_{[2]}$ and $\widetilde{ \Delta} = { \Delta}_{[2]}$ is
defined as
\begin{eqnarray}
{ \Gamma}_{[2]} &=& D_{\mathcal{Y}_{[1]},\;\mathcal{Z}_{[1]}}
{ \Gamma}_{[1]} D_{\mathcal{Y}_{[1]},\;\mathcal{Z}_{[1]}}^{-1}, \\
{ \Delta}_{[2]} &=& D_{\mathcal{Y}_{[1]},\;\mathcal{Z}_{[1]}} {\Delta}_{[1]} D_{\mathcal{Y}_{[1]},\;\mathcal{Z}_{[1]}}^{-1},
\end{eqnarray}
with generic eigenfunction
\begin{eqnarray}
\varphi_{[2]} &:=& D_{\mathcal{Y}_{[1]},\;\mathcal{Z}_{[1]}} (\varphi_{[1]}), \notag \\
&=& \varphi_{[1]} - \mathcal{Y}_{[1]} \Upsilon(\mathcal{Y}_{[1]},\;\mathcal{Z}_{[1]})^{-1}
(I+\lambda I \Omega^{-\dagger}) \Upsilon(\varphi_{[1]},\;\mathcal{Z}_{[1]}).
\end{eqnarray}
and with generic adjoint eigenfunction,
\begin{eqnarray}
\psi_{[2]} &:=& D_{\mathcal{Y}_{[1]},\;\mathcal{Z}_{[1]}}^{-\dag} (\psi_{[1]}), \notag \\
&=& \psi_{[1]} - \mathcal{Z}_{[1]} \Upsilon(\mathcal{Y}_{[1]},\;\mathcal{Z}_{[1]})^{-\dag}
(I + \mu I \Lambda^{-\dagger})\Upsilon(\mathcal{Y}_{[1]},\;\psi_{[1]})^{\dag}.
\end{eqnarray}
For the nth iteration of the binary DT, we choose the eigenfunction $\mathcal{Y}_{[n]}$ that defines the transformation from $\Gamma_{[n]}$, $\Delta_{[n]}$ to $\Gamma_{[n+1]}$, $\Delta_{[n+1]}$. Similarly, we choose the eigenfunction $\mathcal{Z}_{[n]}$ for the adjoint transformation from $\Gamma_{[n]}^{\dagger}$, $\Delta_{[n]}^{\dagger}$ to $\Gamma_{[n+1]}^{\dagger}$, $\Delta_{[n+1]}^{\dagger}$. The Lax operators $\Gamma_{[n]}$ and $\Delta_{[n]}$ exhibit covariance under the binary Darboux transformation.
\begin{equation}
D_{\mathcal{Y}_{[n]},\;\mathcal{Z}_{[n]}} = I -
\mathcal{Y}_{[n]}\Upsilon(\mathcal{Y}_{[n]},\;\mathcal{Z}_{[n]})^{-1} \Omega^{-\dag}
\mathcal{Z}_{[n]}^{\dagger}, \label{BDT1n}
\end{equation}
and adjoint binary DT
\begin{equation}
D_{\mathcal{Y}_{[n]},\;\mathcal{Z}_{[n]}}^{-\dag} = I -
\mathcal{Z}_{[n]}\Upsilon(\mathcal{Y}_{[n]},\;\mathcal{Z}_{[n]})^{-1} \Lambda^{-\dag}
\mathcal{Y}_{[n]}^{\dagger}. \label{BDT2n}
\end{equation}
The transformed operators
\begin{eqnarray}
{ \Gamma}_{[n+1]} &=& D_{\mathcal{Y}_{[n]},\;\mathcal{Z}_{[n]}}
{ \Gamma}_{[n]} D_{\mathcal{Y}_{[n]},\;\mathcal{Z}_{[n]}}^{-1}, \\
{ \Delta}_{[n+1]} &=& D_{\mathcal{Y}_{[n]},\;\mathcal{Z}_{[n]}} {
\Delta}_{[n]} D_{\mathcal{Y}_{[n]},\;\mathcal{Z}_{[n]}}^{-1},
\end{eqnarray}
have generic eigenfunction
\begin{equation}
\varphi_{[n+1]} = \varphi_{[n]} - \mathcal{Y}_{[n]}
\Upsilon(\mathcal{Y}_{[n]},\;\mathcal{Z}_{[n]})^{-1} ( I + \lambda I \Omega^{-\dag})
\Upsilon(\varphi_{[n]},\;\mathcal{Z}_{[n]}),
\end{equation}
and generic adjoint eigenfunction;
\begin{equation}
\psi_{[n+1]} = \psi_{[n]} - \mathcal{Z}_{[n]}
\Upsilon(\mathcal{Y}_{[n]},\;\mathcal{Z}_{[n]})^{-1} ( I + \mu I \Lambda^{-\dag})
\Upsilon(\mathcal{Y}_{[n]},\;\psi_{[n]})^{\dagger}.
\end{equation}
By introducing the matrices $\Xi = (\mathcal{Y}_{1},\;...,\; \mathcal{Y}_{n})$ and $Z =
(\mathcal{Z}_{1},\;...,\; \mathcal{Z}_{n})$, we can represent these findings in the framework of quasi-Gramians, yielding the following expression:
\begin{equation}
\varphi_{[n+1]} = \left\vert
\begin{array}{cc}
  \Upsilon(\Xi,\;Z) & (I + \lambda I \hat{\Omega}^{-\dagger})\Upsilon(\varphi,\;Z) \\
  \Xi & \fbox{$\varphi$}
\end{array}%
\right\vert, \quad \psi_{[n+1]} = \left\vert
\begin{array}{cc}
  \Upsilon(\Xi,\;Z)^{\dag} & (I + \mu I \hat{\Lambda}^{-\dagger}) \Upsilon(\Xi,\;\psi)^{\dag} \\
  Z & \fbox{$\psi$}
\end{array}%
\right\vert, \label{psiBn}
\end{equation}
where $\hat{\Omega} =
\text{diag}(\Omega,\;...,\;\Omega),\;\hat{\Lambda} =
\text{diag}(\Lambda,\;...,\;\Lambda)$, with both $\Omega$ and $\Lambda$ being 2 × 2 matrices.

\section{Quasi-Gramian solutions}
In this
section, we now determine the effect of binary DT $D_{\mathcal{Y},\; \mathcal{Z}}
= I - \xi \Upsilon(\mathcal{Y},\; \mathcal{Z})^{-1} \Omega^{-\dag} \mathcal{Z}^{\dag}$ on
the Lax operator $ \Gamma$, with $\mathcal{Y}_{1},\;...,\;\mathcal{Y}_{n}$ being
eigenfunctions of $ \Gamma$. Similarly, let $\mathcal{Z}_{1},\;...,\;\mathcal{Z}_{n}$ denote the eigenfunctions of the adjoint Lax operator ${ \Gamma}^{\dag}$. The same results apply to the operators ${ \Delta}$ and ${ \Delta}^{\dag}$.

As the binary DT $D_{\mathcal{Y}_{[1]},\;\mathcal{Z}_{[1]}} = D_{\mathcal{Y},\;\mathcal{Z}}$ combines two ordinary DTs, $D_{\mathcal{Y}_{[1]}} = D_{\mathcal{Y}}$ and $D_{\hat{\mathcal{Y}}_{[1]}} = D_{\hat{\mathcal{Y}}}$, it follows that the Lax operator $ \Gamma$ is transformed into a new Lax operator $\widehat{ \Gamma}$ under the binary DT, given by:
\begin{equation}
\widehat{\Gamma}  = {D_{\widehat{\mathcal{Y}}}} {\Gamma}
{D_{\widehat{\mathcal{Y}}}^{-1}},
\end{equation}
we get
\begin{equation}
\widehat{\mathcal{U}} = \mathcal{U} + [\mathcal{J},\; \mathcal{Y} \Upsilon(\mathcal{Y},\; \mathcal{Z})^{-1} \mathcal{Z}^{\dag}].
\label{BDR1}
\end{equation}
Since $\mathcal{U}=[\mathcal{F},\;\mathcal{J}]$, so that
\begin{equation}
\widehat{\mathcal{F}} = \mathcal{F} - \mathcal{Y} \Upsilon(\mathcal{Y},\; \mathcal{Z})^{-1} \mathcal{Z}^{\dag}.
\end{equation}
After $n$ iterations of applying the binary DT $D_{\mathcal{Y},\; \mathcal{Z}}$, the resulting expression is given by:
\begin{eqnarray}
\mathcal{F}_{[n+1]} = \mathcal{F}_{[n]} - \mathcal{Y}_{[n]}
\Upsilon(\mathcal{Y}_{[n]},\;\mathcal{Z}_{[n]})^{-1} \mathcal{Z}_{[n]}^{\dag}= \mathcal{F} - \sum_{i=1}^{n} \mathcal{Y}_{[i]}
\Upsilon(\mathcal{Y}_{[i]},\;\mathcal{Z}_{[i]})^{-1} \mathcal{Z}_{[i]}^{\dag}.
\label{UnBDT}
\end{eqnarray}
Using the notation $\mathcal{F}_{[1]}=\mathcal{F},\;\mathcal{F}_{[2]}=\widehat{\mathcal{F}},\; \mathcal{Y}_{[1]} = \mathcal{Y}_{1}$ and $\mathcal{Z}_{[1]} = \mathcal{Z}_{1}$, and introducing the matrices $\Xi = (\mathcal{Y}_{1},\;...,\; \mathcal{Y}_{n})$ and $Z = (\mathcal{Z}_{1},\;...,\; \mathcal{Z}_{n})$, the result (\ref{UnBDT}) can be expressed in the form of quasi-Gramian as follows:
\begin{equation}
\mathcal{F}_{[n+1]} = \mathcal{F} + \left\vert
\begin{array}{cc}
  \Upsilon(\Xi,\; Z) & Z^{\dag} \\
  \Xi & \fbox{$0_{2}$}
\end{array}
\right\vert, \label{Un123}
\end{equation}
It is worth noting that each $\mathcal{Y}_{i},\; \mathcal{Z}_{i},\;i=1,\;...,\;n$ is a $2\times
2$ matrix. Given that our system of the nc-HNLS equation is formulated in terms of non-commutative objects $u,\;u^{\dag}$, we find it more appropriate to express the quasi-Gramian solution (\ref{Un123}) in terms of these objects. Thus, we introduce the matrices $\mathcal{Y}_{i}$ following a similar approach as in the quasi-Wronskian case. We also define $Z=\Xi Q^{\dag}$, where $Q$ represents a constant matrix of size $2n \times 2n$ and $^{\dag}$ denotes the Hermitian conjugate. It is noted that $\Xi$ and $Z$ adhere to the same dispersion relation and remain unchanged when multiplied by a constant matrix. Consequently, the quasi-Gramian solution (\ref{Un123}) can also be represented as follows:
\begin{equation}
\mathcal{F}_{[n+1]} = \mathcal{F} + \left(\begin{array}{cc}
                                    \left\vert
\begin{array}{cc}
\Upsilon(\Xi,\; Z) & Q \varphi^{\dag} \\
\varphi & \fbox{$ 0 $}%
\end{array}%
\right\vert & \left\vert
\begin{array}{cc}
\Upsilon(\Xi,\; Z) & Q \chi^{\dag} \\
\varphi & \fbox{$ 0 $}%
\end{array}%
\right\vert
\\ \\
\left\vert
\begin{array}{cc}
\Upsilon(\Xi,\; Z) & Q \varphi^{\dag} \\
\chi & \fbox{$ 0 $}%
\end{array}%
\right\vert & \left\vert
\begin{array}{cc}
\Upsilon(\Xi,\; Z) & Q \chi^{\dag} \\
\chi & \fbox{$ 0 $}%
\end{array}%
\right\vert
\end{array}
\right),
\end{equation}
where $\varphi = (\varphi_{1},\;...,\; \varphi_{n})$ and $\chi =
(\chi_{1},\;...,\; \chi_{n})$ are row vectors. Thus, quasi-Gramian
expression for  $u$ and $u^{\dag}$ are given by
\begin{eqnarray}
&&u_{[n+1]} = u + 2 \dot\imath \left\vert
\begin{array}{cc}
  \Upsilon(\Xi,\; Z) & Q \chi^{\dag} \\
  \varphi & \fbox{$0$}
\end{array}
\right\vert, \quad u_{[n+1]}^{\dag} = u^{\dag} + 2 \dot\imath
\left\vert
\begin{array}{cc}
  \Upsilon(\Xi,\; Z) & Q \varphi^{\dag} \\
  \chi & \fbox{$0$}
\end{array}
\right\vert. \label{u2n}
\end{eqnarray}
Equation (\ref{u2n}) represents the quasi-Gramian solutions for the nc-HNLS equation (\ref{EOM}). If we relax the non-commutativity condition, the equation can be simplified and expressed as a ratio of simple Gramians. In the limit of commutativity, we obtain the following expression
\begin{eqnarray}\label{gramm}
u_{[n+1]} = u - 2 \dot\imath \frac{\left\vert
\begin{array}{cc}
  \Upsilon(\Xi,\; Z) & Q \chi^{\dag} \\
  \phi & 0
\end{array}
\right\vert}{\left\vert\Delta(\Xi,\; Z)\right\vert}, \quad
\bar{u}_{[n+1]} = \bar{u} - 2 \dot\imath \frac{\left\vert
\begin{array}{cc}
  \Upsilon(\Xi,\; Z) & Q \varphi^{\dag} \\
  \chi & 0
\end{array}
\right\vert}{\left\vert\Delta(\Xi,\; Z)\right\vert}.
\end{eqnarray}
These expressions define the Gramian solutions of the higher-order NLS equation.

\section{Explicit solutions}
\begin{figure}[H]
        \centering
        \begin{subfigure}{0.23\textwidth}
                \includegraphics[width=\textwidth]{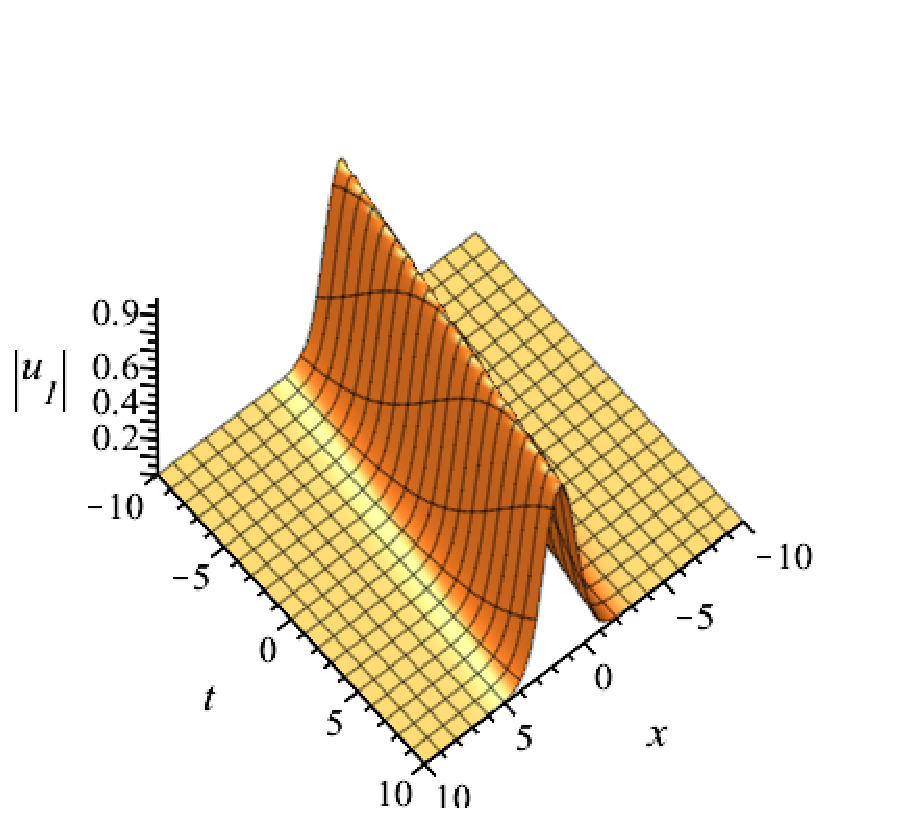}
                \subcaption{}
                \label{fig:fig1}
        \end{subfigure}
        \begin{subfigure}{0.23\textwidth}
                \includegraphics[width=\textwidth]{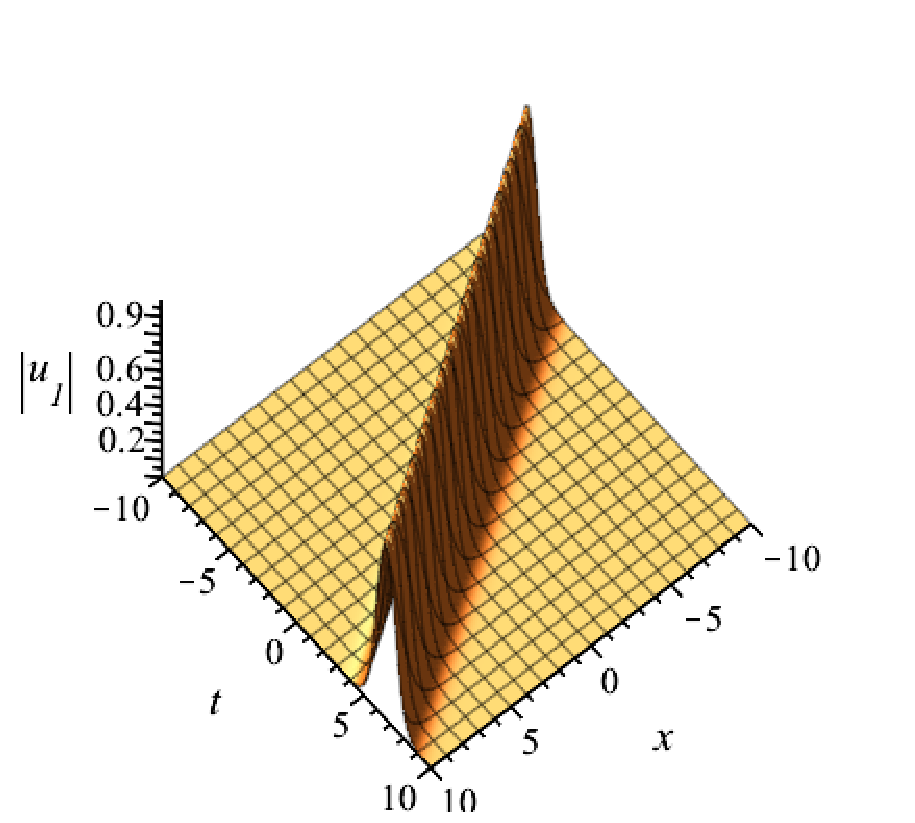}
                \subcaption{}
                \label{fig:fig2}
        \end{subfigure}        
        \caption{Evolution of the solution (\ref{q111111}) with parameters $\alpha_{1} = 1.5,\; \alpha_{2} = \gamma=1,\;c_1 = 0.5,\;q_{1}=2,\;q_{2}=-1,\;(a) \; \lambda = 0.1+0.5\dot\imath,\; (b) \; \lambda = 0.5 \dot\imath$.}
        \label{fig:figure1}
\end{figure}

When $u=0$, spectral problem (\ref{lax1}) and (\ref{lax2}) has the  solution
\begin{equation}
\varphi_{j}= e^{\dot\imath\zeta_{j}} ,\;\quad \chi_{j}=
e^{-\dot\imath\zeta_{j}} , \quad \zeta_{j}=-\lambda_{j} x +
2 \lambda_{j}^{2} (4 \gamma  \,\lambda_{j}^{2}-2 \lambda_{j}  \alpha_{1}-\alpha_{2})t,
\label{phiseed}
\end{equation}
To simplify the notation and work with only $\varphi_1, \ldots, \varphi_n$ and $\chi_1, \ldots, \chi_n$, we introduce the following relabeling. We redefine $\varphi_i$ as $\varphi_{\frac{i+1}{2}}$ for odd values of $i$ (i.e., $i = 1, \ldots, 2n-1$), and set $\varphi_i = 0$ for even values of $i$ (i.e., $i = 2, 4, \ldots, 2n$). Similarly, we relabel $\chi_i$ as $\chi_{\frac{i}{2}}$ for even values of $i$, and $\chi_i = 0$ for odd values of $i$. We then have
\begin{equation}
\varphi = (\varphi_{1},\;0,\;\varphi_{2},\;0,\;
...,\;\varphi_{n},\;0), \quad \chi = (0,\;\chi_{1},\;0,\;
\chi_{2},\; ...,\;0,\; \chi_{n}), \label{phichi}
\end{equation}
Using the notation $\mathcal{Y}_i = \text{diag}(\varphi_i, \chi_i)$ for $i = 1, \ldots, n$, where $\varphi_i$ and $\chi_i$ are given in (\ref{phiseed}), we now focus on the commutative case. In this case, we can express the Gramian solution (\ref{gramm}) as follows.
\begin{equation}
u_{[n+1]} = - 2 \dot\imath \frac{\left\vert
\begin{array}{cc}
  \Upsilon(\Xi,\; Z) & Q \chi^{\dag} \\
  \varphi & 0
\end{array}
\right\vert}{\left\vert\Upsilon(\Xi,\; Z)\right\vert} = -2 \dot\imath
\frac{\mathcal G}{\mathcal R}, \; \text{say} \label{gramm1}
\end{equation}
where $\mathcal R = \text{det}(\mathcal{W})$ and
\begin{equation}\label{W1}
\mathcal{W} = \Upsilon (\Xi,\; Z) = \int Z^{\dag} \mathcal{J} \Xi dx + I_{2n},
\end{equation}
Here, $I_{2n}$ is the identity matrix with dimensions $2n \times 2n$. Constructing the matrix $\Xi$ involves arranging the eigenfunctions $\mathcal{Y}_1, \; \mathcal{Y}_2 ,\; ..., \; \mathcal{Y}_{n}$, where each $\mathcal{Y}_{i}$ represents an eigenfunction of the Lax operators $\Gamma$ and $\Delta$, presented as a $2 \times 2$ matrix. Similarly, assembling the matrix $Z$ involves the eigenfunctions $\mathcal{Z}_1 , \;  \mathcal{Z}_2 ,\;  ...,\;  \mathcal{Z}_{n}$, with $\mathcal{Z}_{i}$ serving as eigenfunctions of the adjoint Lax operators $\Gamma^{\dagger}$ and $\Delta^{\dagger}$. The matrix $\Upsilon(\Xi,\;Z)$ is a matrix $2n \times 2n$, its entries being scalar components $(1 \times 1)$. As we proceed to discuss the non-commutative case, we will consider every component of $\mathcal{Y}_i$ and $\mathcal{Z}_i$ as a matrix. Presenting the matrix $Q$, a constant matrix $2n \times 2n$, we define $Z$ as the outcome of multiplying $\Xi$ by the Hermitian adjoint of $Q$, denoted as $Z = \Xi Q^\dagger$, allowing us to express the equation (\ref{W1}) as
\begin{equation}\label{W2}
\mathcal{W}  = Q \int \Xi^{\dag} \mathcal{J} \Xi dx + I_{2n} = Q \Theta + I_{2n},
\end{equation}
where
\begin{equation}
\Theta = \left(
        \begin{array}{ccccc}
          -\dot\imath \int^{x} \varphi_{1}^{\ast} \varphi_{1} dx & 0_{2} & \hdots & -\dot\imath \int^{x} \varphi_{1}^{\ast} \varphi_{n} dx & 0_{2} \\
          0_{2} & \dot\imath \int^{x} \chi_{1}^{\ast} \chi_{1} dx & \hdots & 0_{2} & \dot\imath \int \chi_{1}^{\ast} \chi_{n} dx \\
          \vdots & \vdots & \ddots & \vdots & \vdots \\
          -\dot\imath \int^{x} \varphi_{n}^{\ast} \varphi_{1} dx & 0_{2} & \hdots & -\dot\imath \int^{x} \varphi_{n}^{\ast} \varphi_{n} dx & 0_{2} \\
          0_{2} & \dot\imath \int^{x} \chi_{n}^{\ast} \chi_{1} dx & \hdots & 0_{2} & \dot\imath \int^{x} \chi_{n}^{\ast} \chi_{n} dx \\
        \end{array}
      \right).
\end{equation}
To derive one soliton ($n=1,\;u_{[2]} \equiv u_{1}$) solution for the commutative Hirota equation. We choose $Q = \left(
                                    \begin{smallmatrix}
                                      q_{1} & q_{2} \\
                                      q_{2} & q_{1}
                                    \end{smallmatrix}
                                  \right)
$, one soliton solution is given by
\begin{equation}\label{q111111}
    u_1 = \frac{8 \,\mathrm{I} c_{1} q_{\mathit{2}} \lambda_{I}^{2} {\mathrm e}^{2 \,\mathrm{I} \xi_1 }  }{4 \,\mathrm{I} \lambda_{I}  c_{1} q_{\mathit{1}} \cosh\! \left(\xi_2 \right)-4 \lambda_{I}^{2} c_{1}^{2}+q_{\mathit{1}}^{2}-q_{\mathit{2}}^{2}}
\end{equation}
where
\begin{eqnarray}
&& \xi_1 = [8 \left( \lambda_R^{4}  - 6  \lambda_{R}^{2}  \lambda_{I}^{2}  +  \lambda_{I}^{4} \right) \gamma + 2 \left(-  \lambda_{R}^{2}  +   \lambda_{I}^{2}  \right) \alpha_{2}  +4 \left(-  \lambda_{R}^{3}  + 3 \lambda_{R}   \lambda_{I} ^{2}  \right) \alpha_{1}]t- \lambda_{R}  x , \\
&& \xi_2 = 8[8 \left(- \lambda_{R}^{3}  +  \lambda_{R}   \lambda_{I}^{2}  \right)   \gamma +  \lambda_{R}  \alpha_{2}+ \left(- \lambda_{I}^{2}  + 3  \lambda_{R}^{2}  \right)   \alpha_{1}] \lambda_{I} t + 2 x \lambda_{I} , \notag
\end{eqnarray}
To visualize this soliton solution, we plot the propagation of the $u_{1}$ soliton in the commutative case with a velocity of $8 \left(- \lambda_{R}^{3}  +  \lambda_{R}   \lambda_{I}^{2}  \right)   \gamma +  \lambda_{R}  \alpha_{2}+ \left(- \lambda_{I}^{2}  + 3  \lambda_{R}^{2}  \right)   \alpha_{1}$, where $\lambda = \lambda_{R} + i \lambda_{I}$. Fig. \ref{fig:figure1} illustrates the behavior of the soliton over time.

\subsection{Noncommutative case} 

We now discuss the noncommutative case.  It has been shown in \cite{goncharenko2001multisoliton} that the behavior of matrix solitons differs from their scalar counterparts. Unlike scalar solitons, which maintain their amplitudes unchanged during interactions, matrix solitons undergo transformations that depend on certain rules. These transformations affect the amplitudes, which are determined by vectors rather than individual values in the noncommutative setting. When considering the case of $n=1$, we choose the solutions $\varphi$ and $\chi$ of the Lax pair to be $2 \times 2$ matrices, given by
\begin{equation}
\varphi_{j}= e^{\dot\imath\zeta_{j}} I_{2} ,\;\quad \chi_{j}=
e^{-\dot\imath\zeta_{j}} I_{2} , \quad \zeta_{j}=-\lambda_{j} x +
2 \lambda_{j}^{2} (4 \gamma  \,\lambda_{j}^{2}-2 \lambda_{j}  \alpha_{1}-\alpha_{2})t,
\label{ncphiseed}
\end{equation}
Each entry in $\varphi$ and $\chi$ and constant matrix $Q$ is a $2 \times 2$ matrix, so that these matrices are given by
\begin{eqnarray}
\varphi = \left(
         \begin{array}{cccc}
           \varphi_{1} & 0 & 0 & 0 \\
           0 & \varphi_{1} & 0 & 0
         \end{array}
       \right), \; \chi = \left(
         \begin{array}{cccc}
           0 & 0 & \chi_{1} & 0 \\
           0 & 0 & 0 & \chi_{1}
         \end{array}
       \right), \; Q = \left(
         \begin{array}{cccc}
           q_{11} & q_{12} & q_{13} & q_{14} \\
           q_{12} & q_{11} & q_{14} & q_{13} \\
           q_{13} & q_{14} & q_{33} & q_{34} \\
           q_{14} & q_{13} & q_{34} & q_{33}
         \end{array}
       \right), \label{phichiH}
\end{eqnarray}
Therefore, the quasi-Gramian expression for $u_{2}$ (which we will now denote as $u^{1}$ for the non-commutative case) can be expressed as follows
\begin{eqnarray}\label{ncq11}
u^{1} &=& {2\dot\imath} \left(\begin{array}{cc}\left\vert
\begin{array}{ccccc}
& & & &   \\
& & \Upsilon(\Xi,\; Z) & & \chi^{11}  \\
& & & &  \\
\varphi_{1} & 0 & 0 & 0 & \fbox{$0$}
\end{array}
\right\vert & \left\vert \begin{array}{ccccc}
& & & &   \\
& & \Upsilon(\Xi,\; Z) & & \chi^{12}  \\
& & & &  \\
\varphi_{1} & 0 & 0 & 0 & \fbox{$0$}
\end{array}
\right\vert \\
\left\vert \begin{array}{ccccc}
& & & &   \\
& & \Upsilon(\Xi,\; Z) & & \chi^{11}  \\
& & & &  \\
0 & \varphi_{1} & 0 & 0  & \fbox{$0$}
\end{array}
\right\vert & \left\vert \begin{array}{ccccc}
& & & &   \\
& & \Upsilon(\Xi,\; Z) & & \chi^{12}  \\
& & & &  \\
0 & \varphi_{1}  &  0 & 0 & \fbox{$0$}
\end{array}
\right\vert
\end{array}%
\right), \notag \\ &=& {2 \dot\imath} \left(\begin{array}{cc}
                                      u_{11} & u_{12} \\
                                      u_{21} & u_{22}
                                    \end{array}
                                    \right), \; \text{say},
\end{eqnarray}
where $\chi^{11} = (q_{13} \chi_{1}^{\ast} \; q_{14} \chi_{1}^{\ast}
\; q_{33} \chi_{1}^{\ast} \; q_{34} \chi_{1}^{\ast})^{\dag},\;
\chi^{12} = (q_{14} \chi_{1}^{\ast} \; q_{13} \chi_{1}^{\ast} \;
q_{34} \chi_{1}^{\ast} \; q_{33} \chi_{1}^{\ast})^{\dag}$ and
$\Upsilon$ is the potential defined in (\ref{W1}) with each entry being a
$2 \times 2$ matrix.

In the context of non-commutative systems, the soliton solution (\ref{ncq11}) is intricately influenced by both the spectral parameter $\lambda$ and the elements composing the matrix $Q$. When specific entries, such as $q_{13}$ and $q_{14}$, are deliberately set to zero, the resulting solutions for $u_{11}$, $u_{12}$, $u_{21}$, and $u_{22}$ appear trivial. And, where $q_{13} = q_{14} = -2$, the graphical representations of solutions $u_{11}$, $u_{12}$, $u_{21}$, and $u_{22}$ manifesting as a single consolidated plot instead of the originally intended four. Noteworthy is the fact that under this condition, all solitons propagate with a consistent amplitude of $0.2163$ units. Unlike this symmetry, when $q_{13} \neq q_{14}$, the resulting graphs exhibit a variety of double- and single-peaked patterns for each component of the matrix $u^{1}$ (see Fig. \ref{fig:figure3}). It is worth highlighting that solitons $u_{12}$ and $u_{21}$ advance with an amplitude of 0.3860 units of large peak and 0.2317 units of small peak, while the solitons $u_{11}$ and $u_{22}$ display amplitude of 1.9797 units. Additionally, an intriguing situation unfolds when we choose $q_{12} = -q_{14} = -2$ and $q_{11} = q_{13} = q_{33} = q_{34} = 0$. Under these conditions, we notice a consistent single-peaked soliton for $u_{11}$ and $u_{22}$ with an amplitude of 0.2214 units, while simultaneously observing a kink pattern in $u_{12}$ and $u_{21}$ of maximum height 0.8858 units (as seen in Fig. \ref{fig:figure4}).

\begin{figure}[H]
        \centering
        \begin{subfigure}{0.32\textwidth}
                \includegraphics[width=\textwidth]{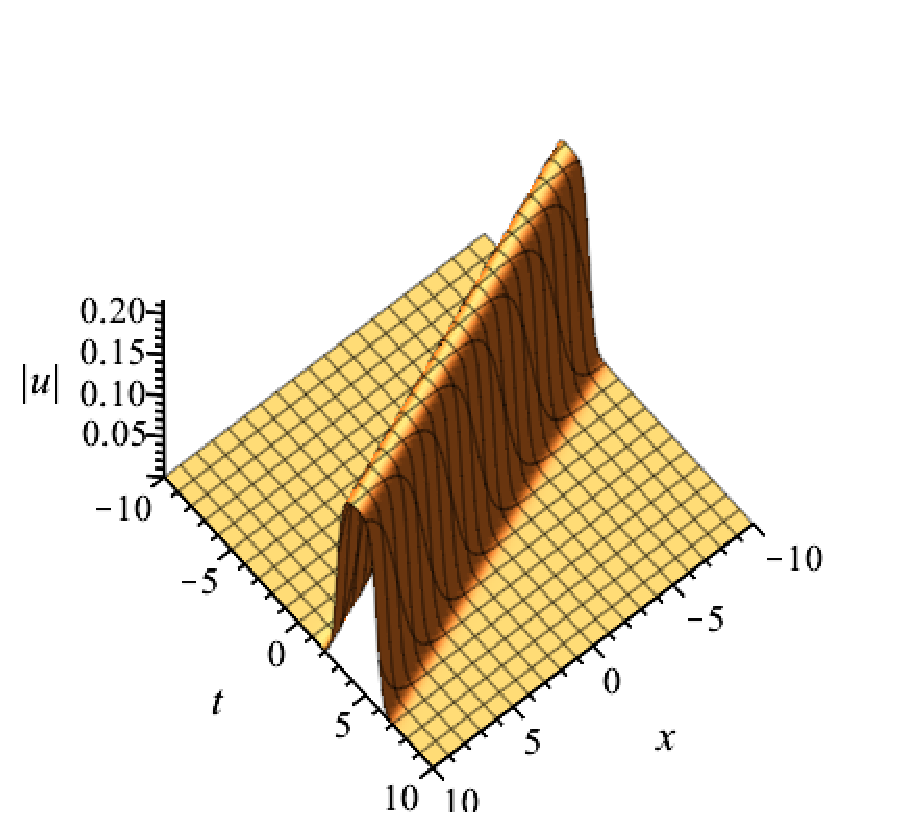}
                \label{fig:q11solitoncase2}
        \end{subfigure}
        \caption{Evolution of the solution (\ref{ncq11}) with parameters $\alpha_{1} = 1.5,\; \alpha_{2} = \gamma=1,\;c_1 = 0.5,\;q_{11}=0.5,\;q_{12}=0,\;q_{13}=q_{14}=-1,\;q_{33}=-0.2,\;q_{34}=-0.1,\; \lambda = -0.1+0.5\dot\imath$.}
        \label{fig:figure2}
\end{figure}

\begin{figure}[H]
        \centering
        \begin{subfigure}{0.23\textwidth}
                \includegraphics[width=\textwidth]{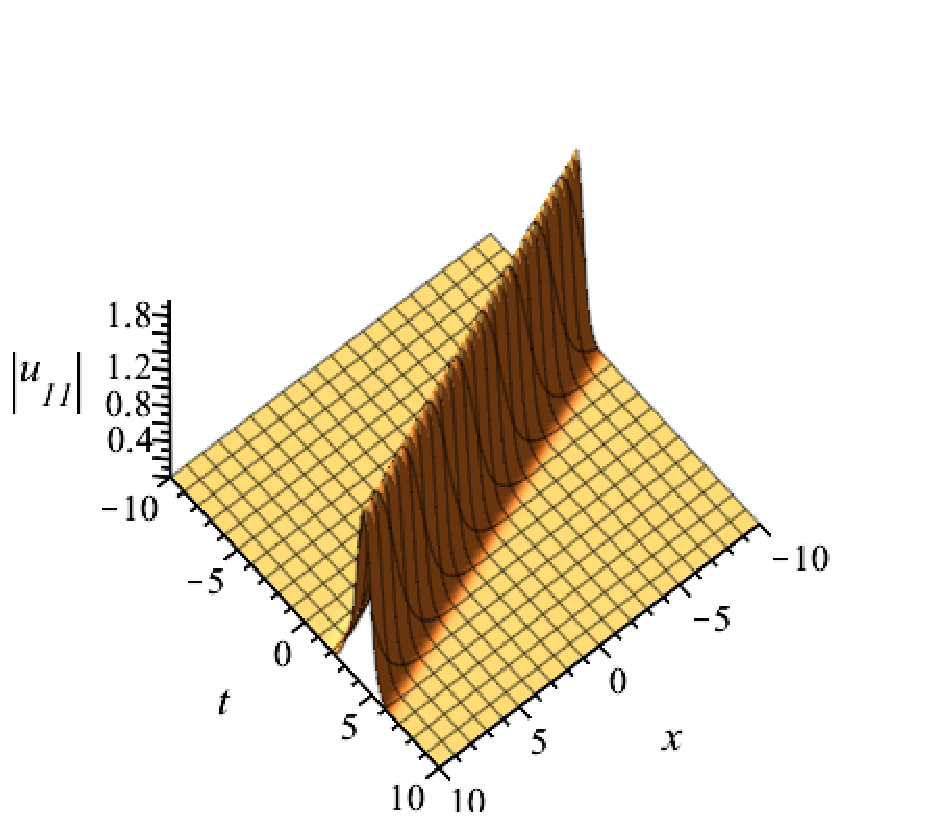}
                \subcaption{}
                 \label{fig:u11solitoncase3}
        \end{subfigure}
        \begin{subfigure}{0.23\textwidth}
                \includegraphics[width=\textwidth]{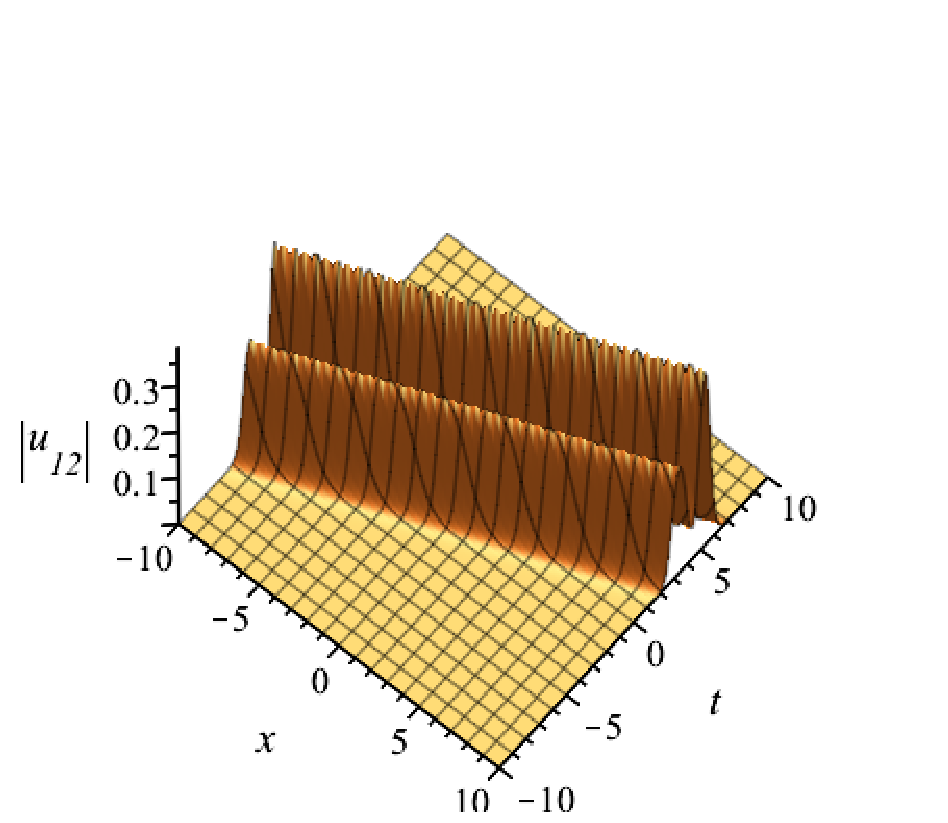}
                \subcaption{}
                \label{fig:u12solitoncase3}
        \end{subfigure}
                \begin{subfigure}{0.23\textwidth}
                \includegraphics[width=\textwidth]{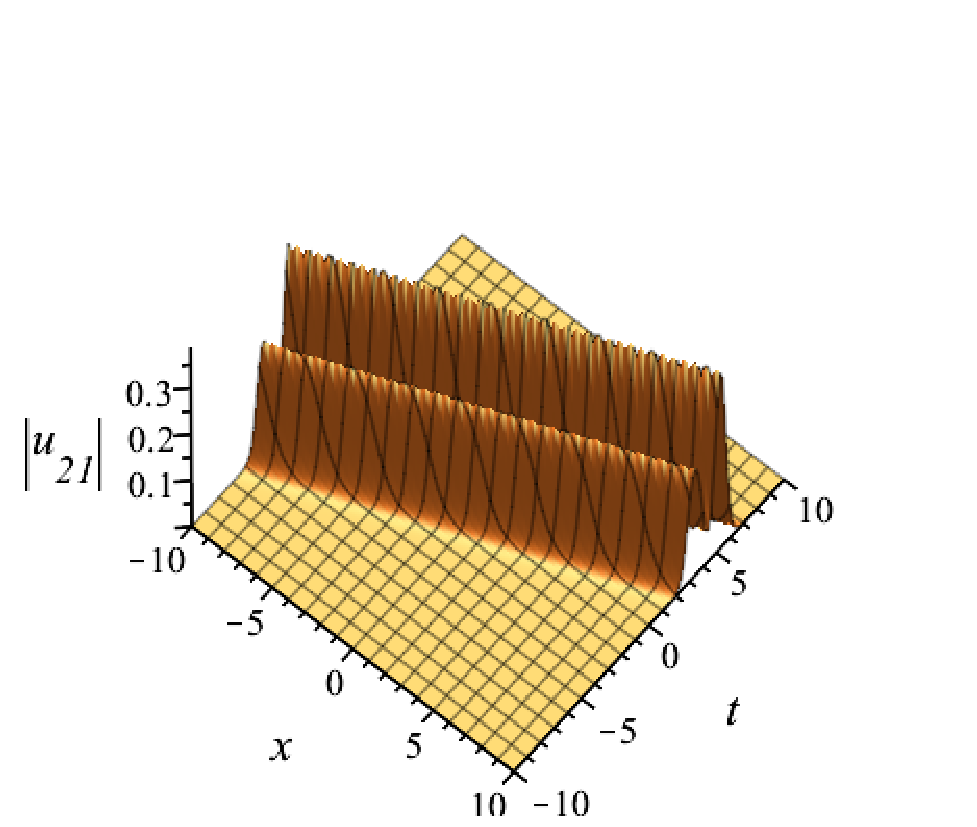}
                \subcaption{}
                \label{fig:u21solitoncase3}
        \end{subfigure}
        \begin{subfigure}{0.23\textwidth}
                \includegraphics[width=\textwidth]{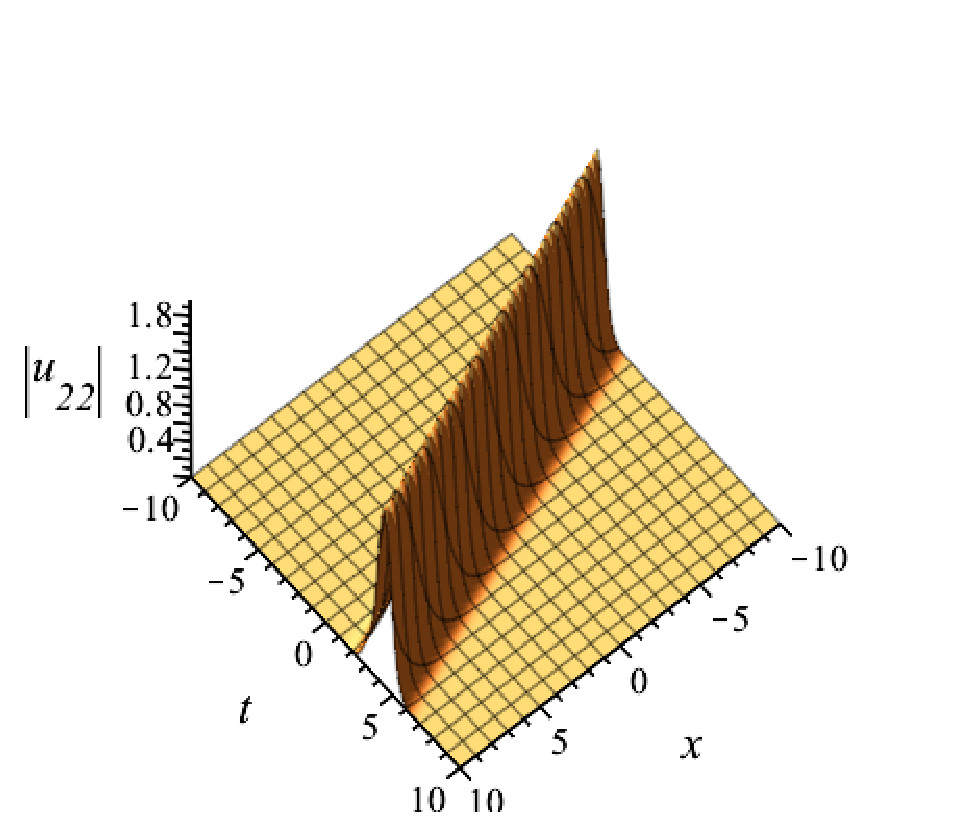}
                \subcaption{}
                \label{fig:u22solitoncase3}
        \end{subfigure}
       \caption{Evolution of the solution (\ref{ncq11}) with parameters $\alpha_{1} = 1.5,\; \alpha_{2} = \gamma=1,\;c_1 = 0.5,\;q_{11}=0.5,\;q_{12}=-0.2,\;q_{13}=-1,\; q_{14}=-0.6,\;q_{33}=0,\;q_{34}=0.1,\; \lambda = -0.1+0.5\dot\imath$.}
        \label{fig:figure3}
\end{figure}

\begin{figure}[H]
        \centering
        \begin{subfigure}{0.23\textwidth}
                \includegraphics[width=\textwidth]{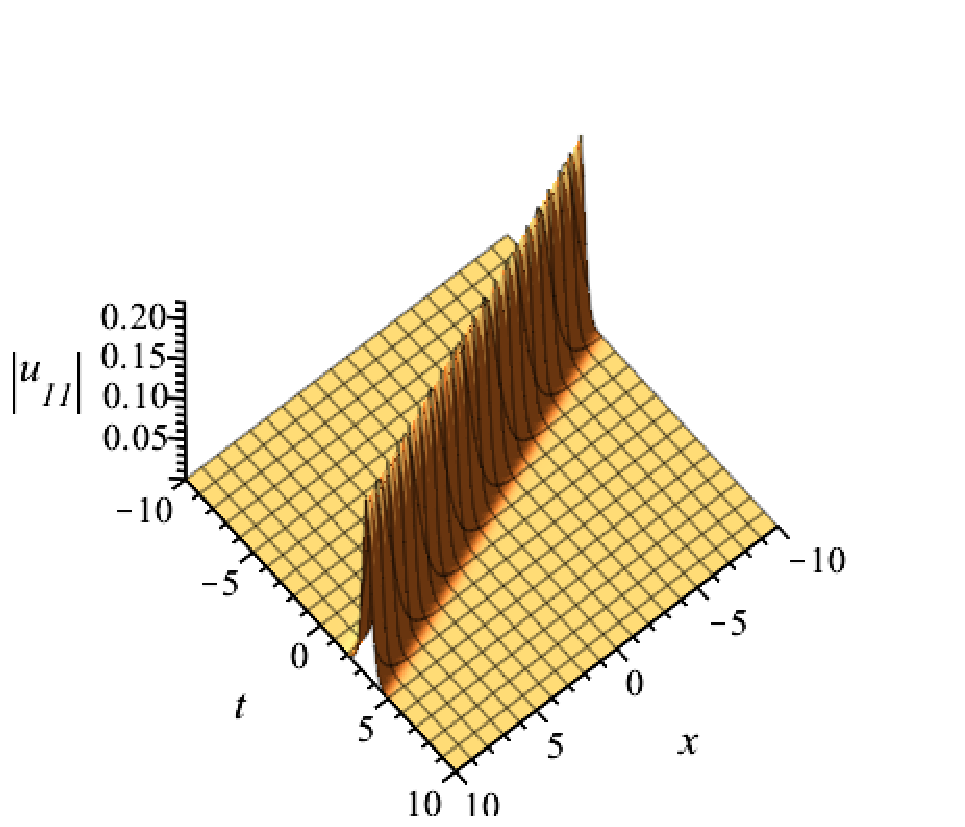}
                \subcaption{}
                 \label{fig:u11solitoncase4}
        \end{subfigure}
        \begin{subfigure}{0.23\textwidth}
                \includegraphics[width=\textwidth]{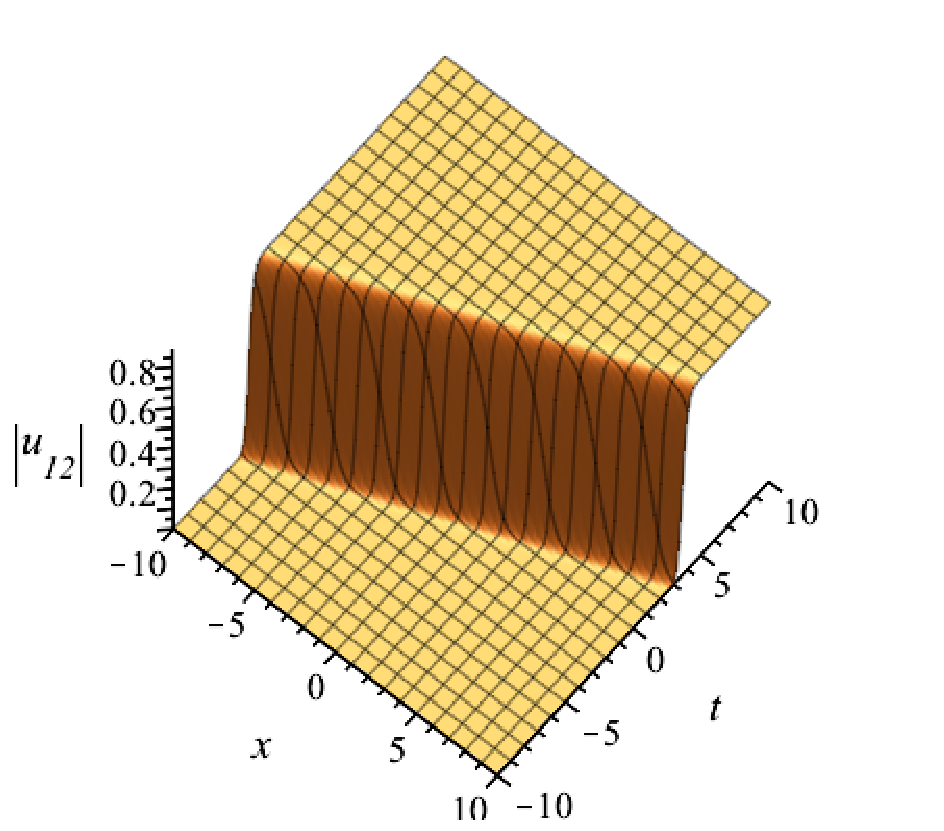}
                \subcaption{}
                \label{fig:u12solitoncase4}
        \end{subfigure}
                \begin{subfigure}{0.23\textwidth}
                \includegraphics[width=\textwidth]{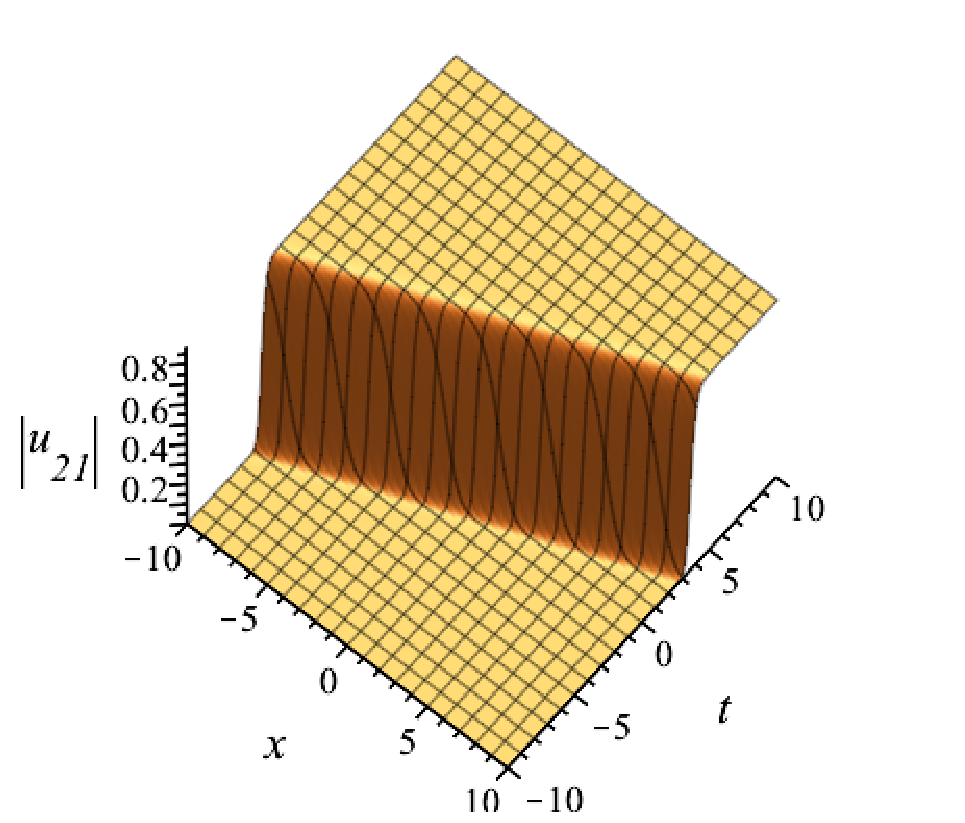}
                \subcaption{}
                \label{fig:u21solitoncase4}
        \end{subfigure}
        \begin{subfigure}{0.23\textwidth}
                \includegraphics[width=\textwidth]{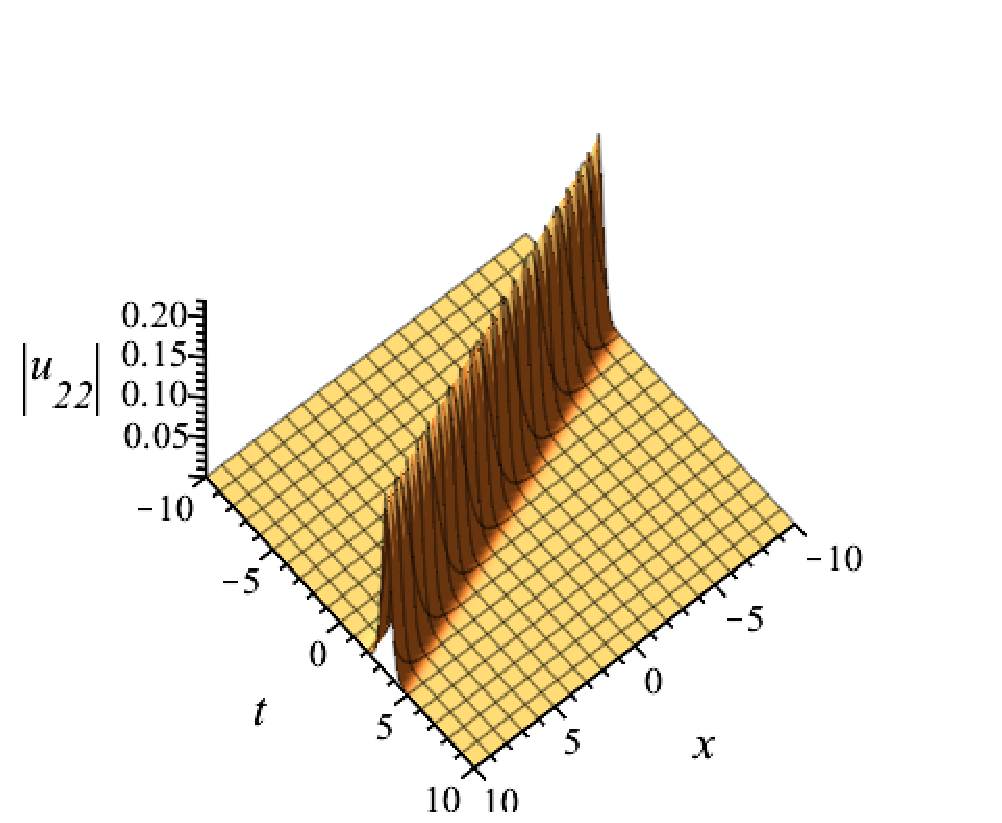}
                \subcaption{}
                \label{fig:u22solitoncase4}
        \end{subfigure}
       \caption{Evolution of the solution (\ref{ncq11}) with parameters $\alpha_{1} = 1.5,\; \alpha_{2} = \gamma=1,\;c_1 = 0.5,\;q_{11}=0,\;q_{12}=-2,\;q_{13}=0,\; q_{14}=2,\;q_{33}=0,\;q_{34}=0,\; \lambda = -0.1+0.5\dot\imath$.}
        \label{fig:figure4}
\end{figure}

The solution (\ref{ncq11}) includes several particular cases. In the case where both $\alpha_2$ and $\gamma$ are zero, the solution (\ref{ncq11}) takes a different form. It becomes a solution of a non-commutative generalization of the complex modified Korteweg-de Vries (KdV) equation and is further reduced to the standard modified KdV equation when the variable $u$ is real-valued. Additionally, if we set $\gamma$ to zero, we obtain the solution of the non-commutative extension of the Hirota equation. On the other hand, when we simultaneously set $\alpha_1$ and $\alpha_2$ to zero, we have the solution of the non-commutative Lakshmanan–Porsezian–Daniel (LPD) equation. Lastly, if we set $\alpha_1$ and $\gamma$ to zero, we get the solution of non-commutative generalization of the nonlinear Schr{\"o}dinger (NLS) equation. 

\begin{figure}[H]
        \centering
        \begin{subfigure}{0.23\textwidth}
                \includegraphics[width=\textwidth]{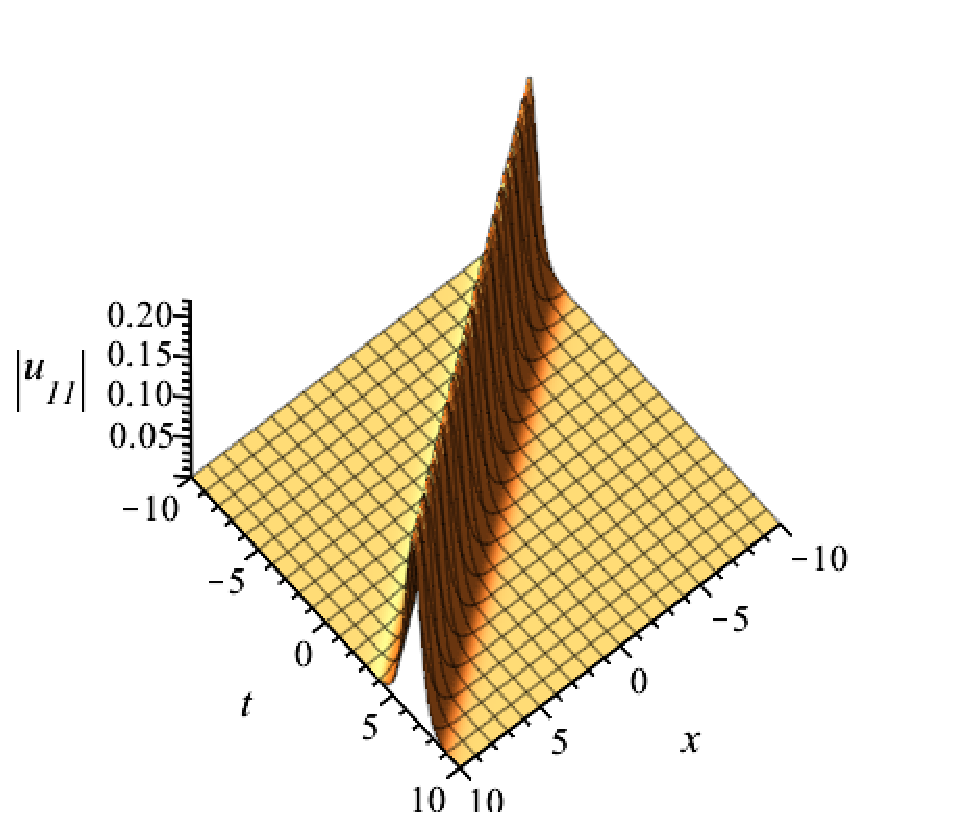}
                \subcaption{}
                 \label{fig:u11solitonmkdv}
        \end{subfigure}
        \begin{subfigure}{0.23\textwidth}
                \includegraphics[width=\textwidth]{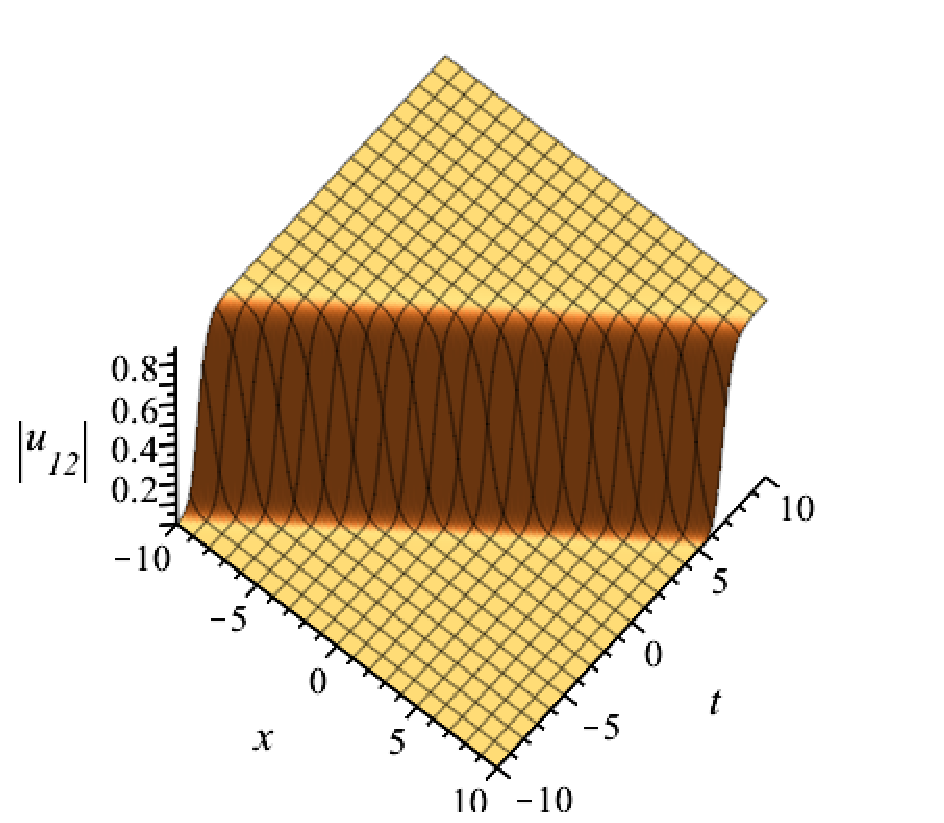}
                \subcaption{}
               \label{fig:u12solitonmkdv}
        \end{subfigure}
                \begin{subfigure}{0.23\textwidth}
                \includegraphics[width=\textwidth]{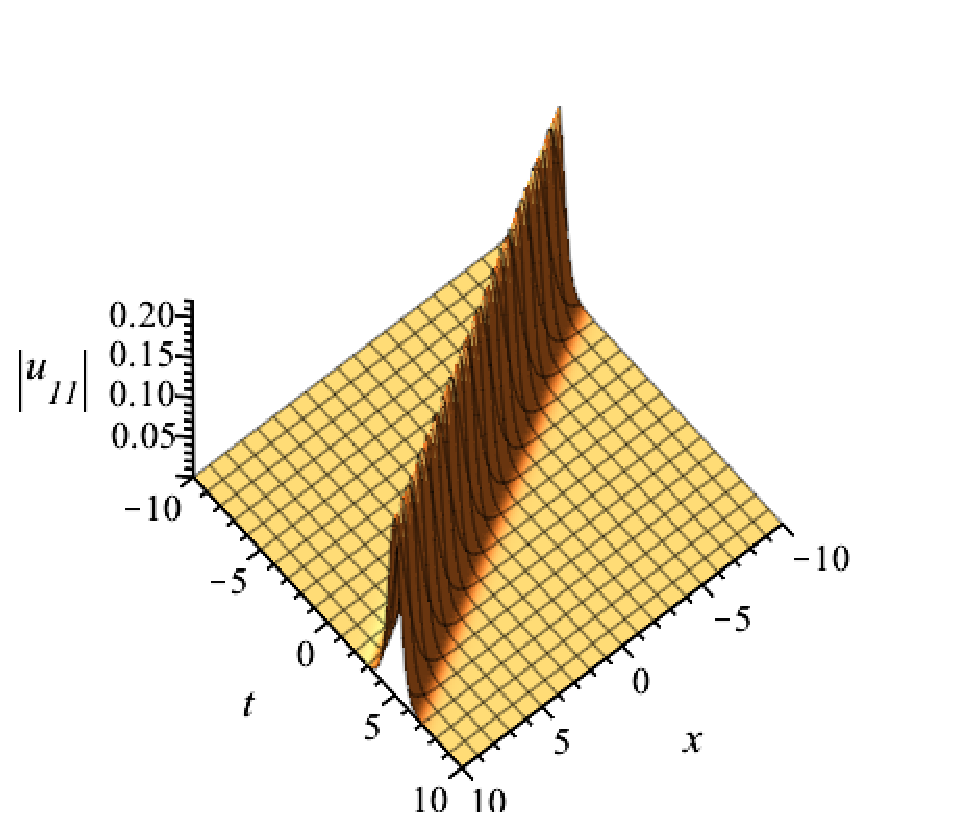}
                \subcaption{}
                \label{fig:u11solitonnHE}
        \end{subfigure}
        \begin{subfigure}{0.23\textwidth}
                \includegraphics[width=\textwidth]{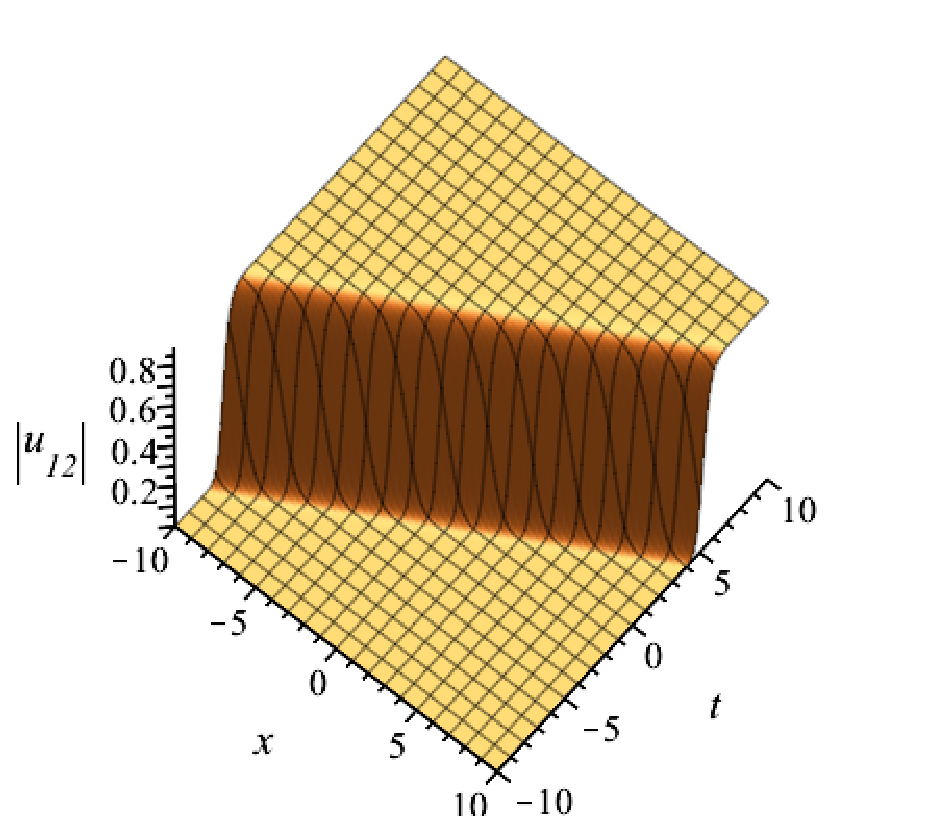}
                \subcaption{}
                \label{fig:u12solitonnHE}
        \end{subfigure}
       \caption{The profiles of $u_{11}$ and $u_{12}$ with $\nu = \alpha_2 = 0$ in (a) and (b) and, with $\nu=0$ in (c) and (d). All other parameters are the same as in Fig. \ref{fig:figure4}.}
        \label{fig:figure5}
\end{figure}

\begin{figure}[H]
        \centering
        \begin{subfigure}{0.23\textwidth}
                \includegraphics[width=\textwidth]{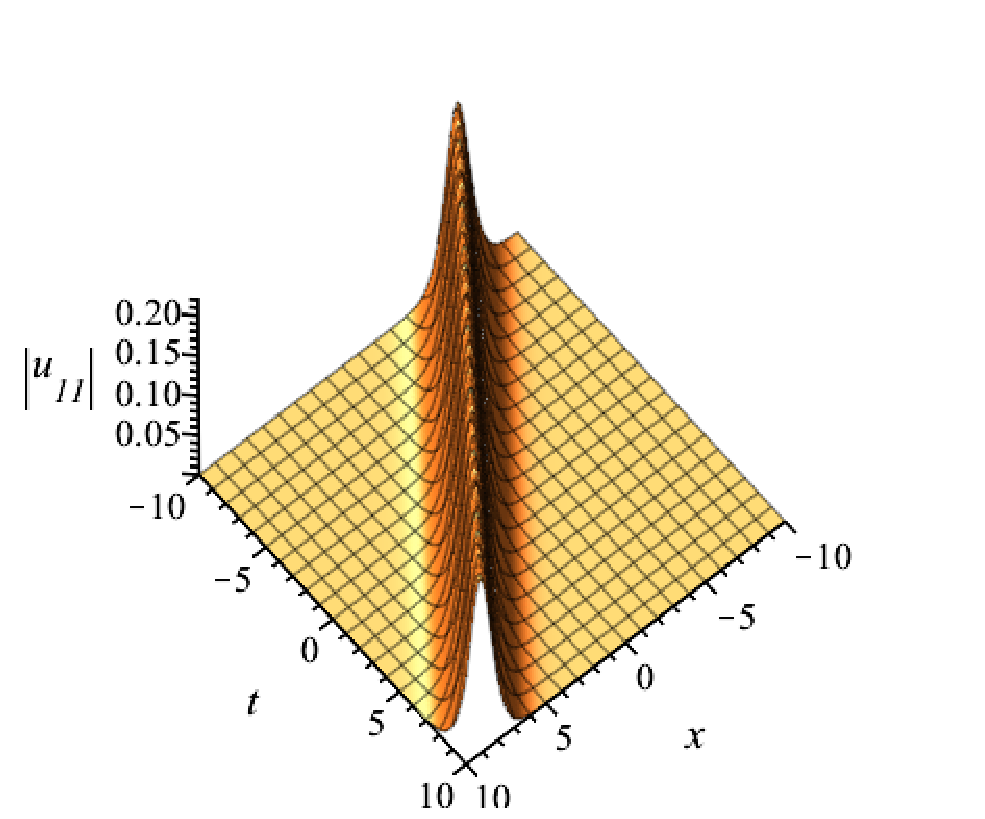}
                \subcaption{}
                 \label{fig:u11solitonLPD}
        \end{subfigure}
        \begin{subfigure}{0.23\textwidth}
                \includegraphics[width=\textwidth]{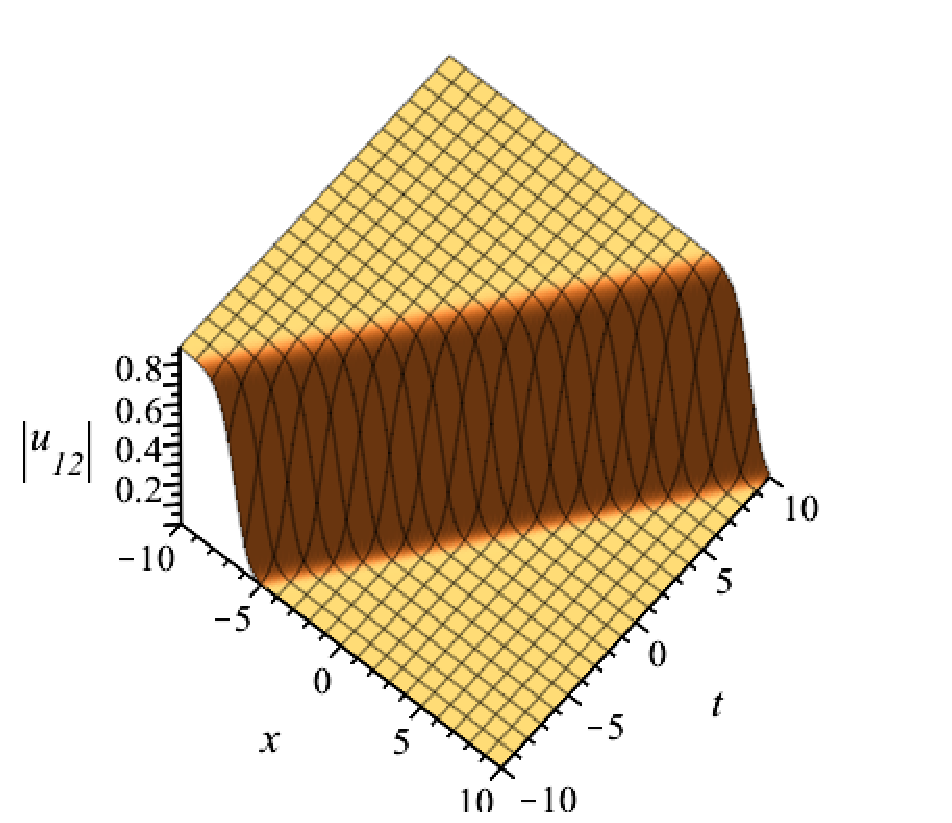}
                \subcaption{}
               \label{fig:u12solitonLPD}
        \end{subfigure}
                \begin{subfigure}{0.23\textwidth}
                \includegraphics[width=\textwidth]{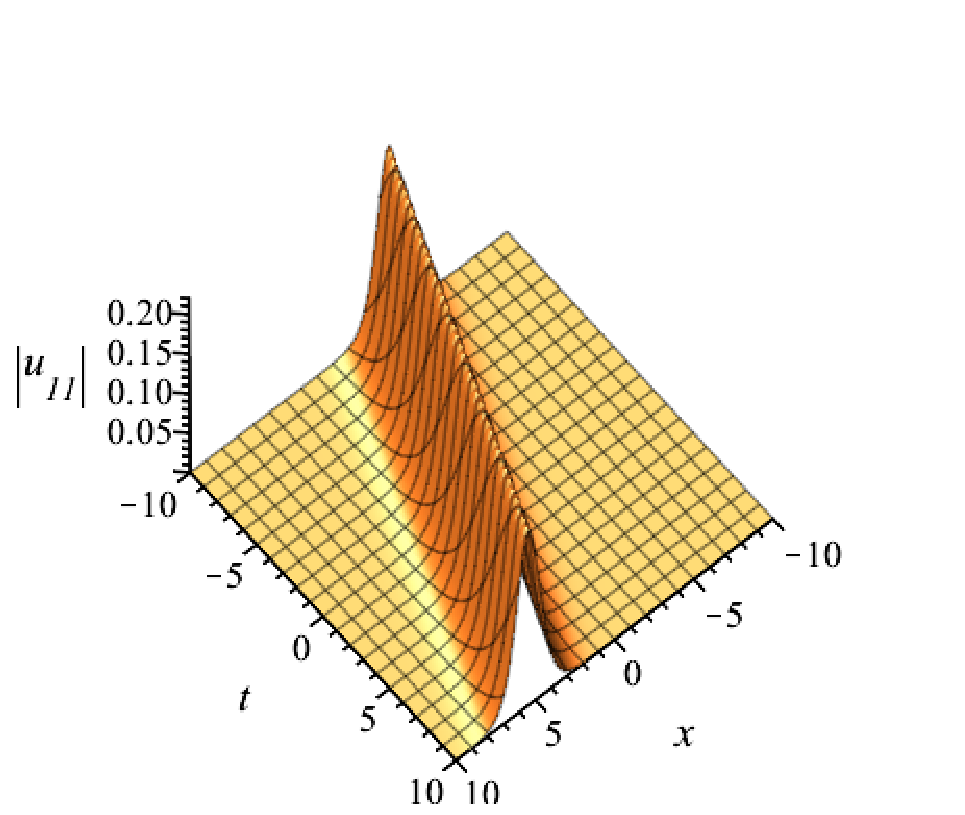}
                \subcaption{}
                \label{fig:u11solitonNLS}
        \end{subfigure}
        \begin{subfigure}{0.23\textwidth}
                \includegraphics[width=\textwidth]{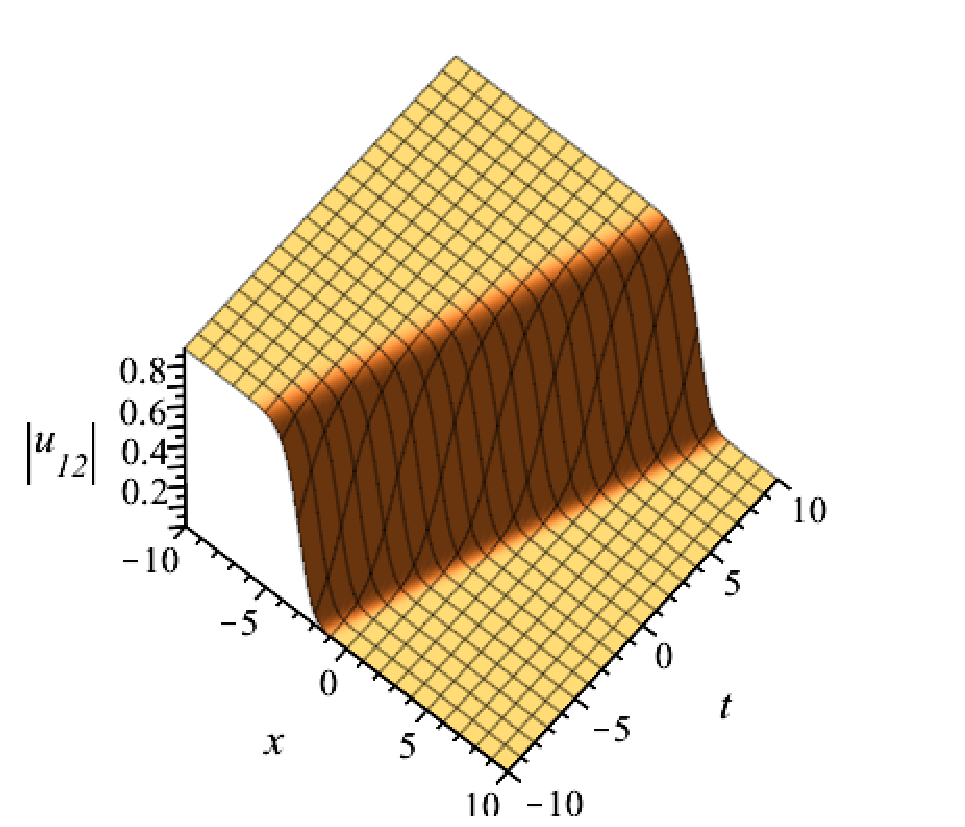}
                \subcaption{}
                \label{fig:u12solitonNLS}
        \end{subfigure}
       \caption{The profiles of $u_{11}$ and $u_{12}$ with $\alpha_1 = \alpha_2 = 0$ in (a) and (b) and, with $\alpha_1 = \nu=0$ in (c) and (d). All other parameters are the same as in Fig. \ref{fig:figure4}.}
        \label{fig:figure6}
\end{figure}

\begin{figure}[H]
        \centering
        \begin{subfigure}{0.23\textwidth}
                \includegraphics[width=\textwidth]{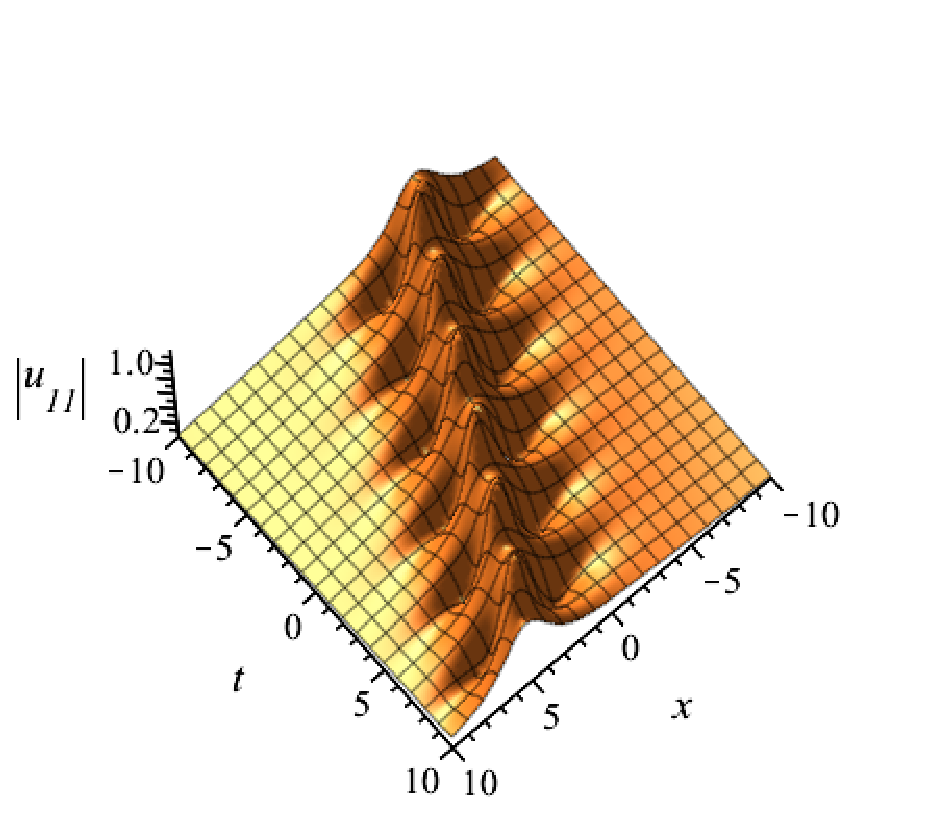}
                \subcaption{}
                 \label{fig:u11breather}
        \end{subfigure}
        \begin{subfigure}{0.23\textwidth}
                \includegraphics[width=\textwidth]{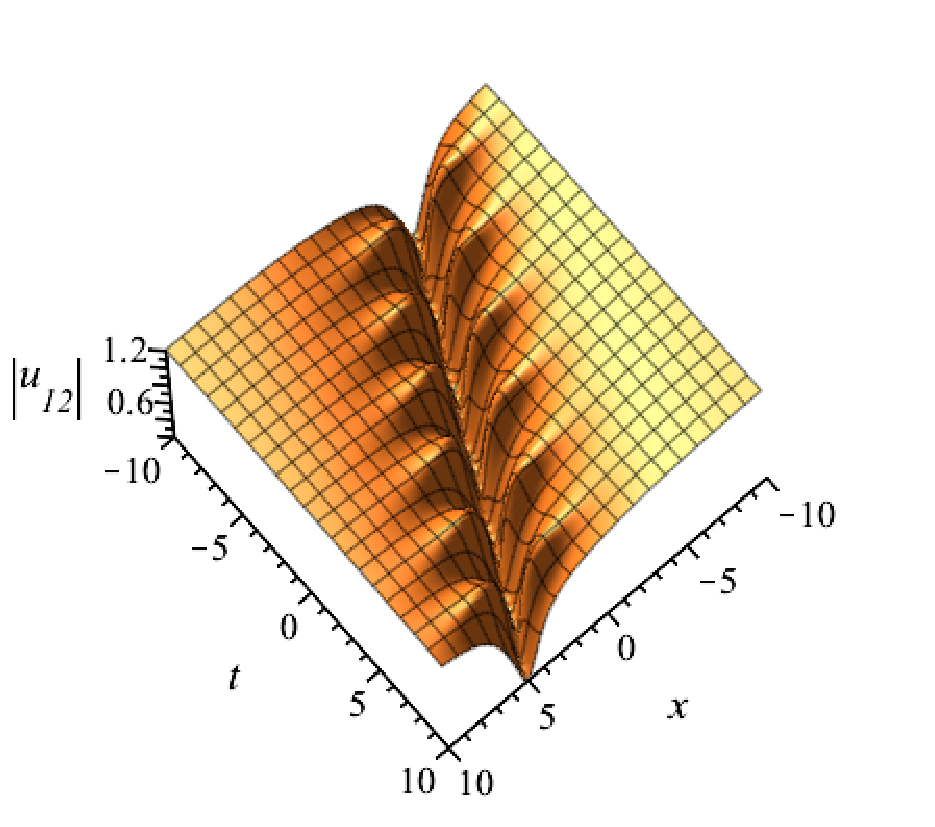}
                \subcaption{}
               \label{fig:u12breather}
        \end{subfigure}
                \begin{subfigure}{0.23\textwidth}
                \includegraphics[width=\textwidth]{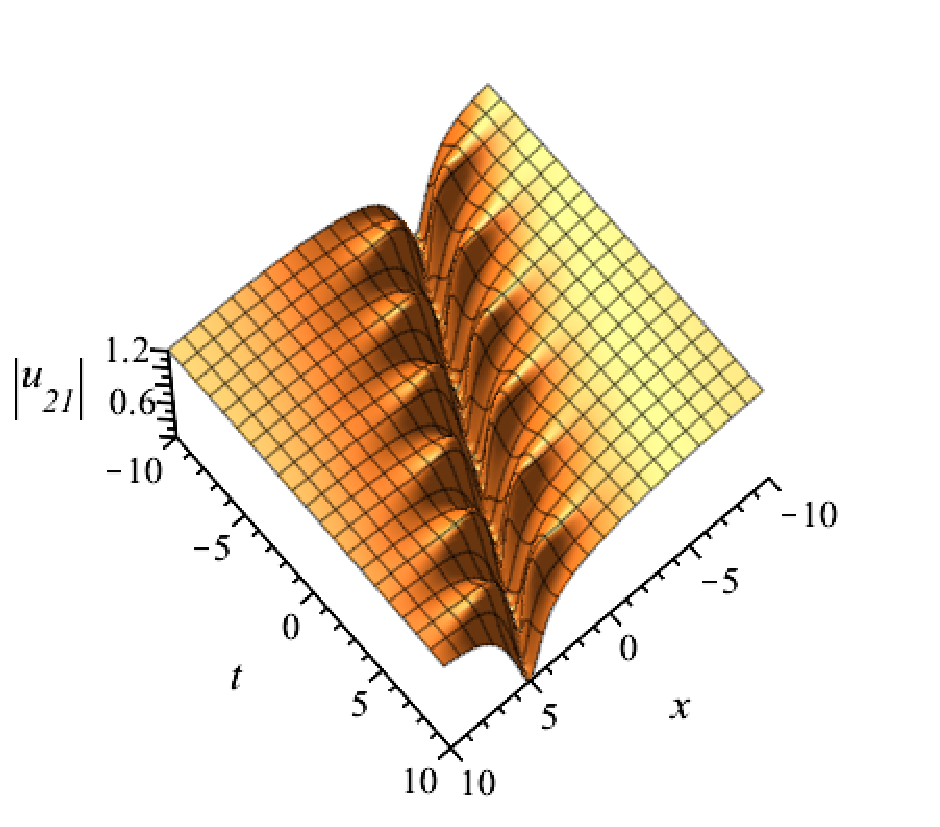}
                \subcaption{}
                \label{fig:u21breather}
        \end{subfigure}
        \begin{subfigure}{0.23\textwidth}
                \includegraphics[width=\textwidth]{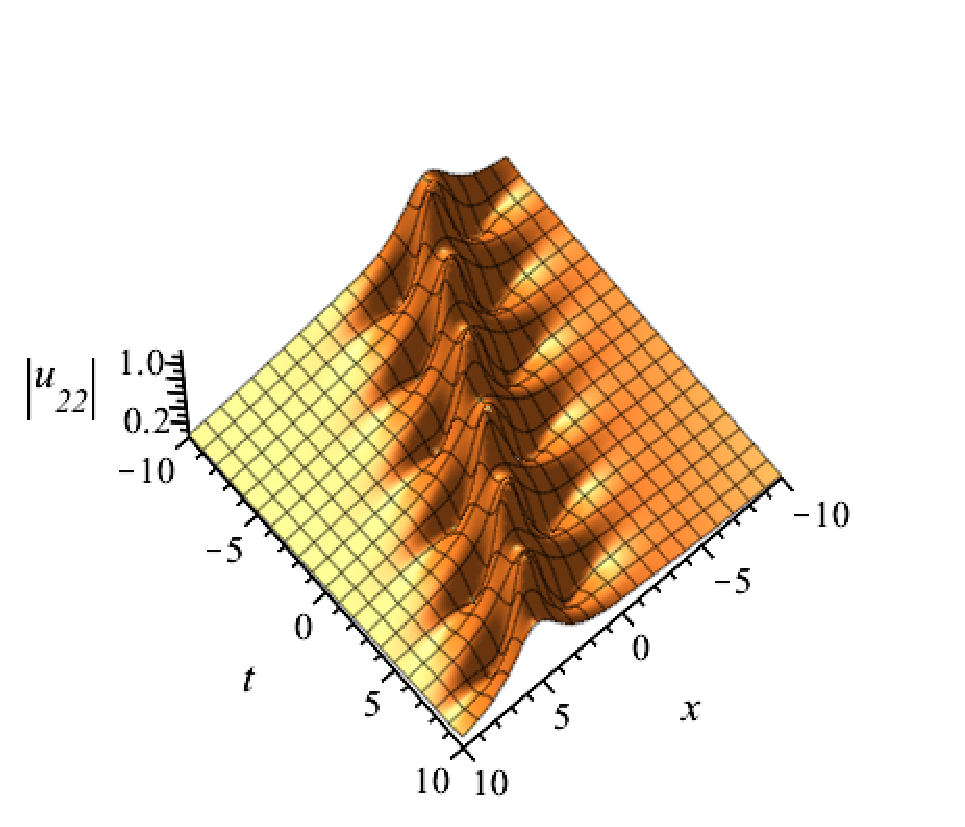}
                \subcaption{}
                \label{fig:u22breather}
        \end{subfigure}
       \caption{The profiles of $u_{ij},\;i,\;j=1,\;2$ with $\alpha_1 = 0.5,\;\alpha_2 = \nu=1,\;\lambda_{1} = 0.5i,\; \lambda_2 
 =-0.1-0.1i$. All other parameters are the same as in Figure \ref{fig:figure4}.}
        \label{fig:figure7}
\end{figure}

\begin{figure}[H]
        \centering
        \begin{subfigure}{0.23\textwidth}
                \includegraphics[width=\textwidth]{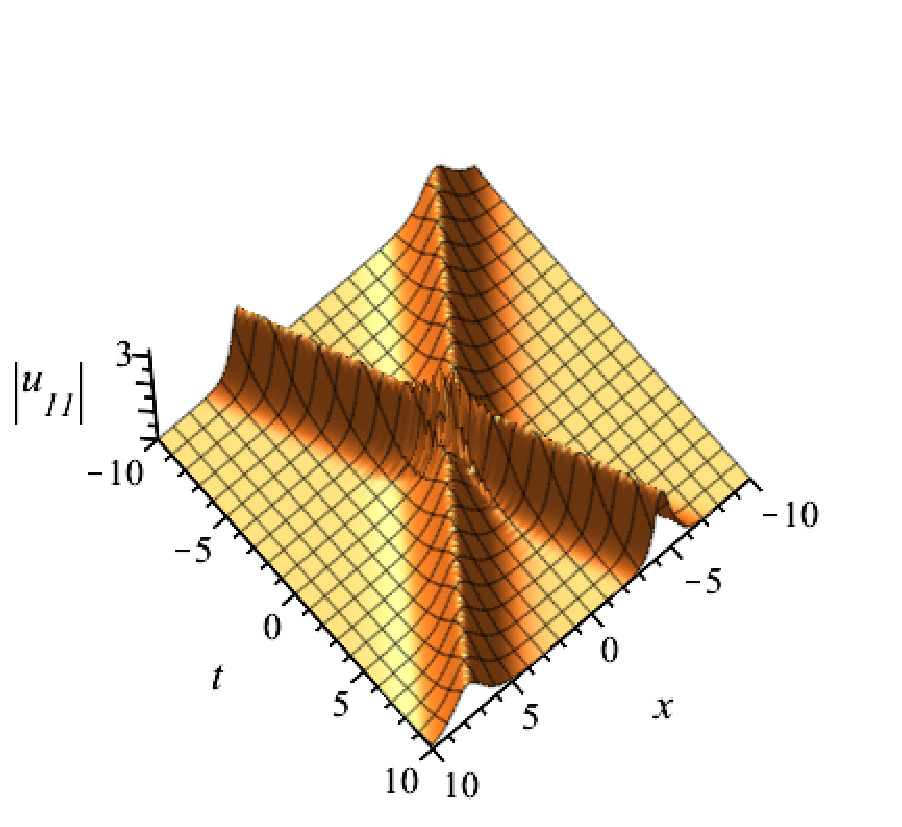}
                \subcaption{}
                 \label{fig:u11int}
        \end{subfigure}
        \begin{subfigure}{0.23\textwidth}
                \includegraphics[width=\textwidth]{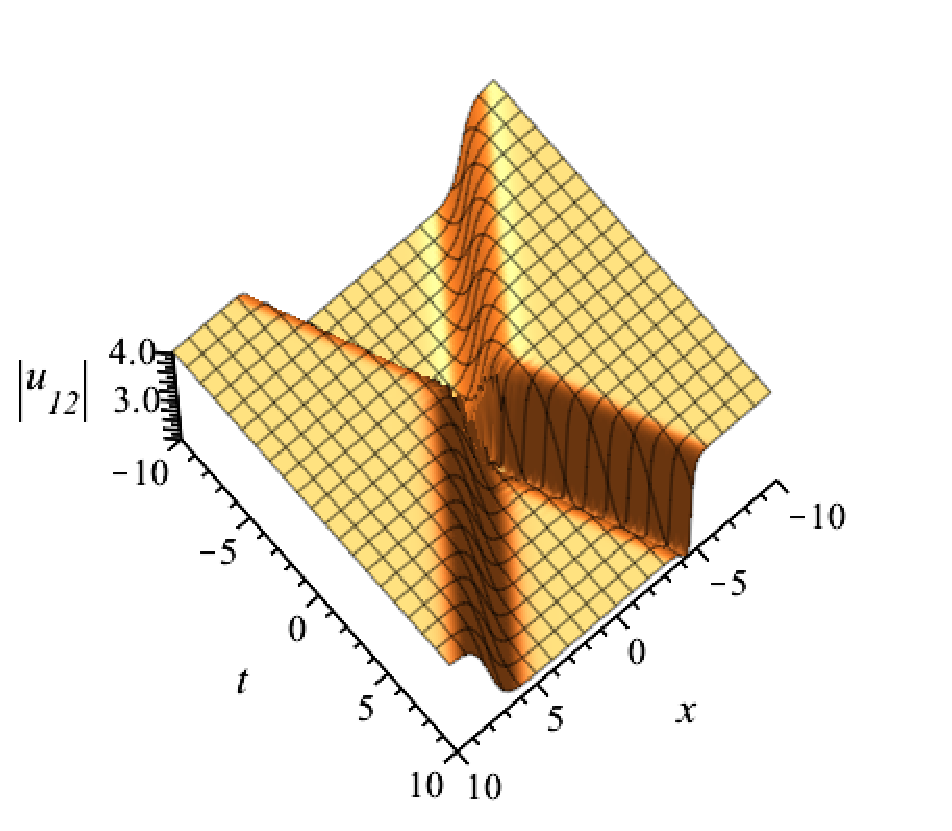}
                \subcaption{}
               \label{fig:u12int}
        \end{subfigure}
                \begin{subfigure}{0.23\textwidth}
                \includegraphics[width=\textwidth]{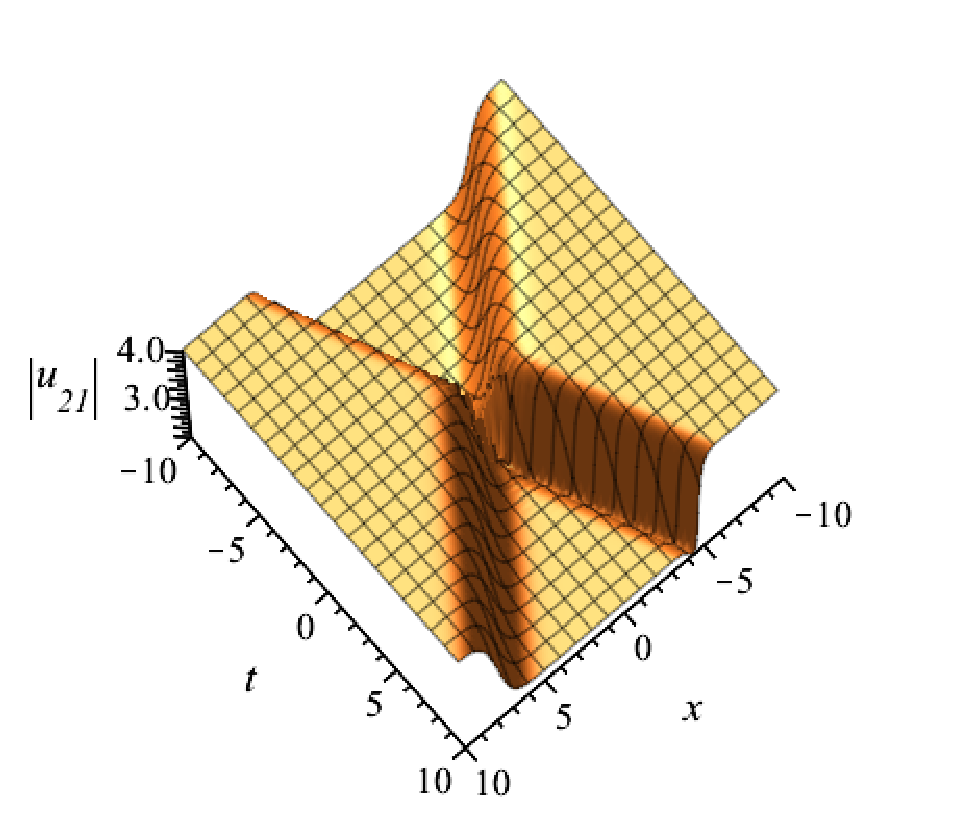}
                \subcaption{}
                \label{fig:u21int}
        \end{subfigure}
        \begin{subfigure}{0.23\textwidth}
                \includegraphics[width=\textwidth]{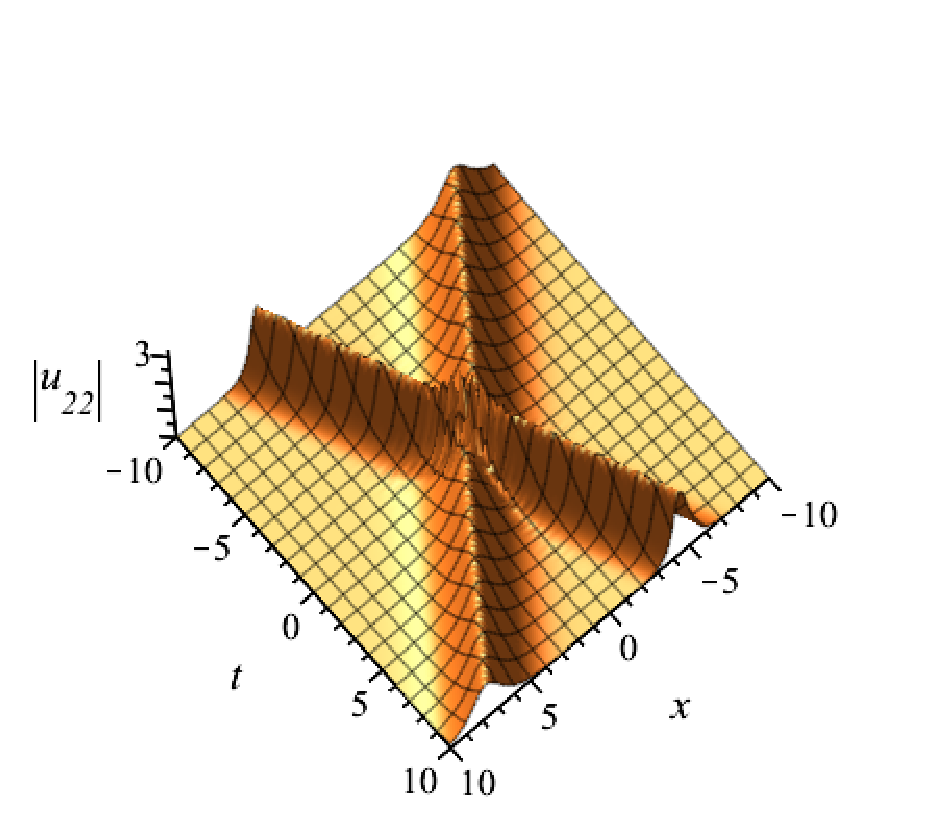}
                \subcaption{}
                \label{fig:u22int}
        \end{subfigure}
       \caption{The profiles of $u_{ij},\;i,\;j=1,\;2$ with $\alpha_1 = 0.5,\;\alpha_2 = \nu=1,\;\lambda_{1} = 0.6i,\; \lambda_2 
 =-1.1-1.1i$.}
        \label{fig:figure8}
\end{figure}

\begin{figure}[H]
        \centering
        \begin{subfigure}{0.23\textwidth}
                \includegraphics[width=\textwidth]{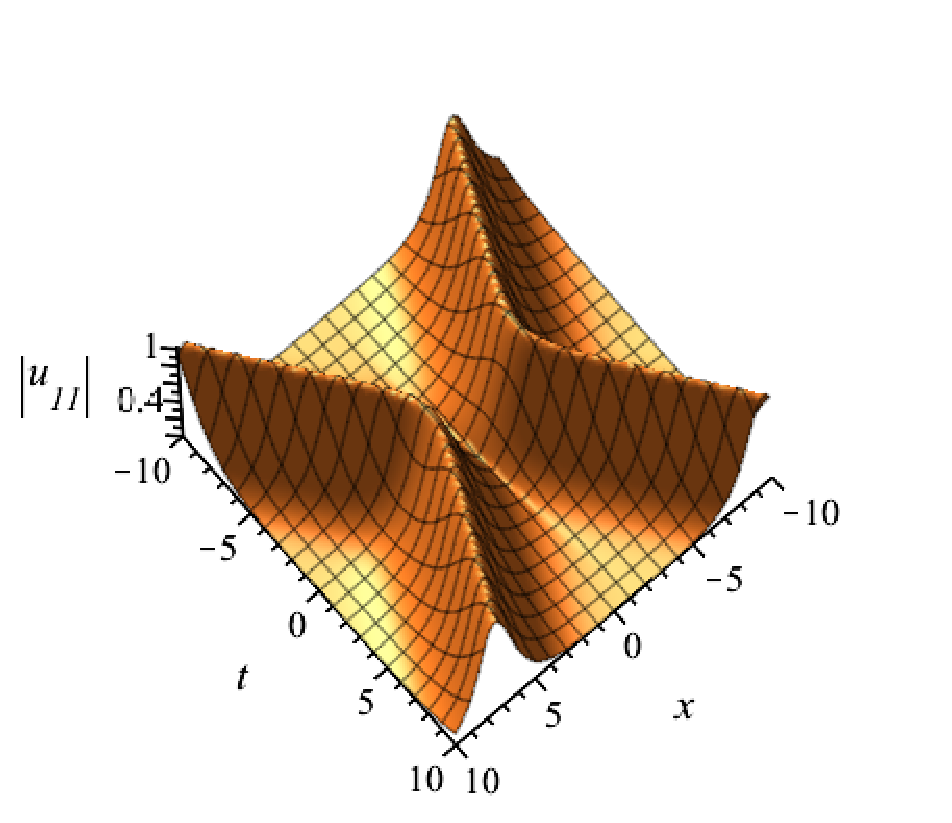}
                \subcaption{}
                 \label{fig:u11pro}
        \end{subfigure}
        \begin{subfigure}{0.23\textwidth}
                \includegraphics[width=\textwidth]{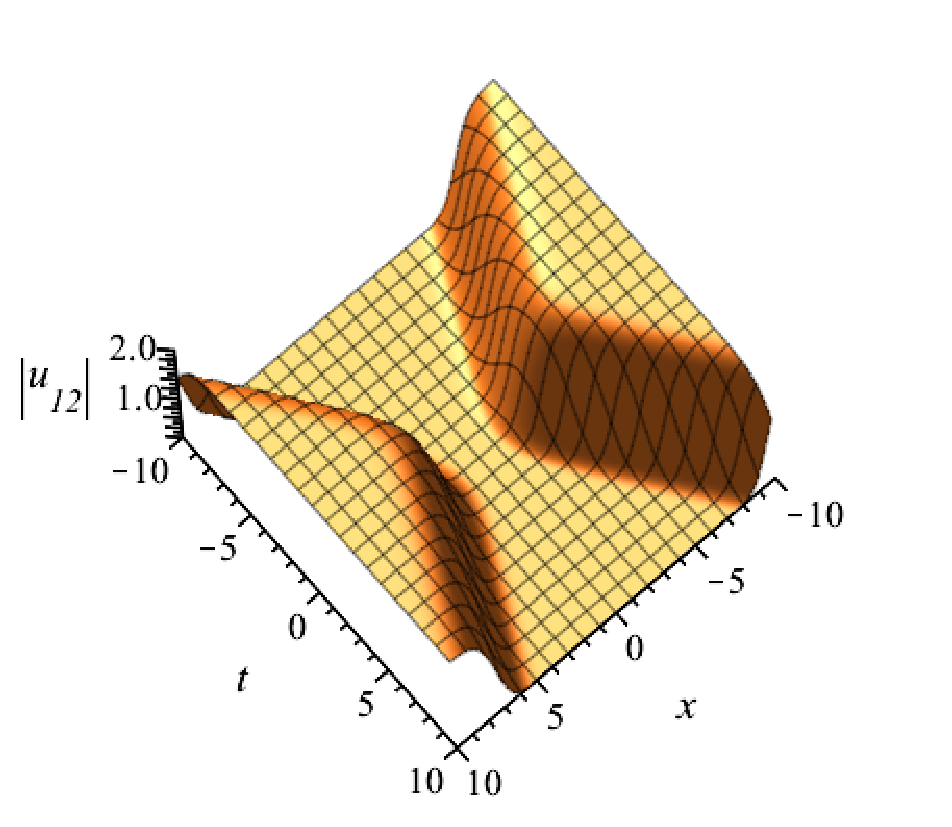}
                \subcaption{}
               \label{fig:u12pro}
        \end{subfigure}
                \begin{subfigure}{0.23\textwidth}
                \includegraphics[width=\textwidth]{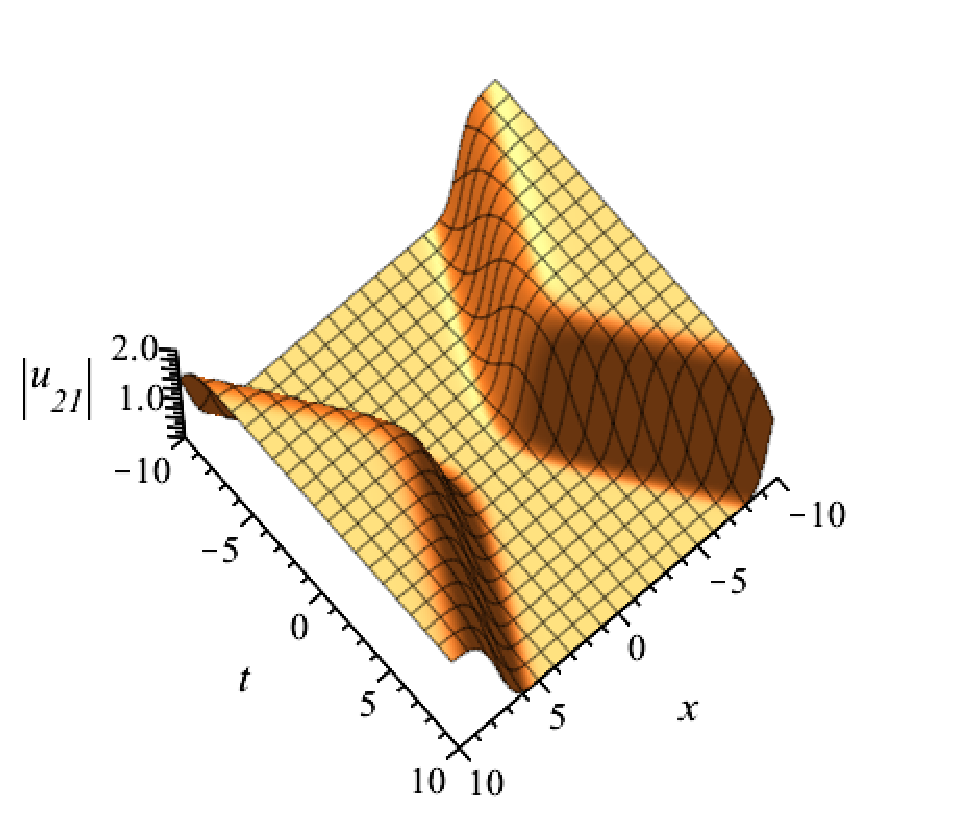}
                \subcaption{}
                \label{fig:u21pro}
        \end{subfigure}
        \begin{subfigure}{0.23\textwidth}
                \includegraphics[width=\textwidth]{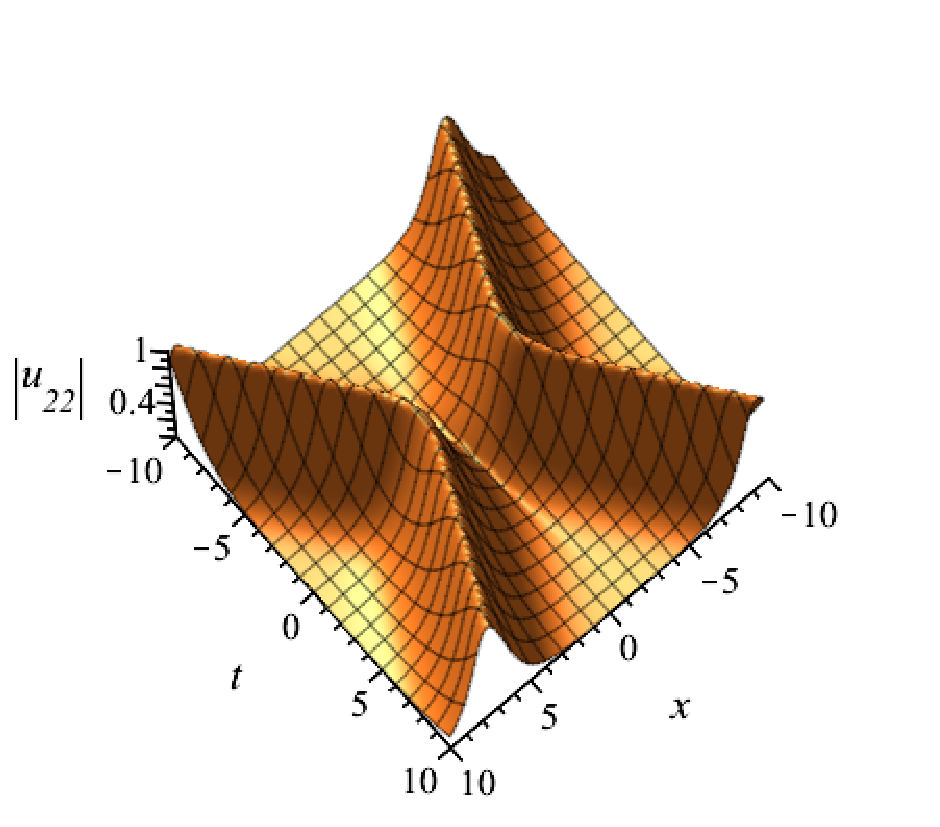}
                \subcaption{}
                \label{fig:u22pro}
        \end{subfigure}
       \caption{The profiles of $u_{ij},\;i,\;j=1,\;2$ with $\alpha_1 = 0.5,\;\alpha_2 = \nu=1,\;\lambda_{1} = 0.1+0.5i,\; \lambda_2  =-0.5i$.}
        \label{fig:figure9}
\end{figure}

In summary, studying the non-commutative version is important because it gives us different choices for arranging solitons. These arrangements depend not only on the spectral parameter $\lambda$ but also on values in a matrix.  Similarly, two soliton solutions for the non-commutative case are depicted in Figs. \ref{fig:figure7}-\ref{fig:figure9}

From Fig. \ref{fig:figure7}, it can be seen that solitons with breather-like structures can be seen in $u_{ij}, i,\;j=1,\;2$ components. This is because when the force between the two solitons is strong enough, bound-state solitons can merge into breather-like solitons. This process is known as soliton fusion, and it occurs when two solitons combine to form a single soliton.

Figs. \ref{fig:figure8}-\ref{fig:figure9} show two-soliton solutions of nc-HNLS equation, which correspond to the interaction of two individual solitons, that is, two distinct lumps of energy moving at different velocities that interact and scatter without changing their shape. Other multisoliton expressions can be obtained by repeatedly applying the Darboux transformation to the seed solution. The three-soliton configuration represents three distinct amplitudes of soliton scattering. Note that, we have omitted the explicit expression of soliton solutions for non-commutative cases as it is long and cumbersome. 

\vspace{6pt}

\large{\textbf{Acknowledgment:}}
We acknowledge the support of the National Natural Science Foundation of China, Nos. 11835011 and 12375006.

\vspace{6pt}

\large{\textbf{Data Availability:}}
Not Applicable.

\section{Concluding remarks}
This study explored the non-commutative extension of the higher-order nonlinear Schr{\"o}dinger equation. Through Darboux and binary Darboux transformations, we expressed in quasi-Wronskian and quasi-Gramian forms. These solutions were intricately linked to the non-commutative higher-order nonlinear schr{\"o}didnger equation and its associated Lax pair. Importantly, we demonstrated single-, double-peaked, kink, and bright solitons in non-commutative settings. Our approach offers a powerful tool not only for understanding the non-commutative higher order nonlinear Schr{\"o}dinger equation but also for constructing multisolitons in related non-commutative integrable systems.


\begin{thebibliography}{999}
     \bibitem{ablowitz2004discrete}
     Ablowitz, M. J., Prinari, B., and Trubatch, A. D. (2004). \textit{Discrete and continuous nonlinear Schr{\"o}dinger systems (Vol. 302).} Cambridge University Press.
     
     \bibitem{kivshar2003optical}
     Kivshar, Y. S., and Agrawal, G. P. (2003). \textit{Optical solitons: from fibers to photonic crystals.} Academic press.
     
     \bibitem{agrawal2000nonlinear}
     Agrawal, G. P. (2000). Nonlinear fiber optics. \textit{In Nonlinear Science at the Dawn of the 21st Century} (pp. 195-211). Berlin, Heidelberg: Springer Berlin Heidelberg.
     
     \bibitem{daniel1995davydov}
     Daniel, M., and Deepamala, K. (1995). Davydov soliton in alpha helical proteins: higher order and discreteness effects. \textit{Physica A: Statistical Mechanics and its Applications}, 221(1-3), 241-255.
     
     \bibitem{lakshmanan1988effect}
     Lakshmanan, M., Porsezian, K., and Daniel, M. (1988). Effect of discreteness on the continuum limit of the Heisenberg spin chain. \textit{Physics Letters A}, 133(9), 483-488.
     
     \bibitem{porsezian1992integrability}
     Porsezian, K., Daniel, M., andLakshmanan, M. (1992). On the integrability aspects of the one‐dimensional classical continuum isotropic biquadratic Heisenberg spin chain. \textit{Journal of mathematical physics}, 33(5), 1807-1816.
     
     \bibitem{ankiewicz2014extended}
     Ankiewicz, A., Wang, Y., Wabnitz, S., and Akhmediev, N. (2014). Extended nonlinear Schrödinger equation with higher-order odd and even terms and its rogue wave solutions. \textit{Physical Review E}, 89(1), 012907.
     
     \bibitem{ankiewicz2014higher}
     Ankiewicz, A., and Akhmediev, N. (2014). Higher-order integrable evolution equation and its soliton solutions. \textit{Physics Letters A}, 378(4), 358-361.
     
     \bibitem{nimmo2000applications}
     Nimmo, J. J. C., Gilson, C. R., and Ohta, Y. (2000). Applications of Darboux transformations to the self-dual Yang-Mills equations. \textit{Theoretical and Mathematical Physics}, 122(2), 239-246.
     
     \bibitem{goncharenko2001multisoliton}
     Goncharenko, V. M. (2001). Multisoliton solutions of the matrix KdV equation. \textit{Theoretical and Mathematical Physics}, 126(1), 81-91.
     
     \bibitem{lakshmanan1977continuum}
     Lakshmanan, M. (1977). Continuum spin system as an exactly solvable dynamical system. \textit{Physics Letters A}, 61(1), 53-54.
     
     \bibitem{hirota1973exact}
     Hirota, R. (1973). Exact envelope‐soliton solutions of a nonlinear wave equation. \textit{Journal of Mathematical Physics}, 14(7), 805-809.
     
     \bibitem{minwalla2000noncommutative} 
     Minwalla, S., Van Raamsdonk, M., and Seiberg, N. (2000). Noncommutative perturbative dynamics. \textit{Journal of High Energy Physics}, 2000(02), 020.
     
     \bibitem{furuta2000ultraviolet}
     Furuta, K., and Inami, T. (2000). Ultraviolet property of noncommutative Wess–Zumino–Witten model. \textit{Modern Physics Letters A}, 15(15), 997-1002.
     
     \bibitem{Seiberg1999}
     Seiberg, N., and Witten, E. (1999). String theory and noncommutative geometry. \textit{Journal of High Energy Physics}, 1999(09), 032.
     
     \bibitem{dimakis2000bicomplexes}
     Dimakis, A., and Muller-Hoissen, F. (2000). Bicomplexes, integrable models, and noncommutative geometry. \textit{arXiv preprint hep-th/0006005}.
     
     \bibitem{lechtenfeld2001noncommutative}
     Lechtenfeld, O., and Popov, A. D. (2001). Noncommutative multi-solitons in 2+ 1 dimensions. \textit{Journal of High Energy Physics}, 2001(11), 040.
     
     \bibitem{Lechtenfeld2005}
     Lechtenfeld, O., Mazzanti, L., Penati, S., Popov, A. D., and Tamassia, L. (2005). Integrable noncommutative sine-Gordon model. \textit{Nuclear Physics B}, 705(3), 477-503.
     
     \bibitem{Gilson2007} 
     Gilson, C. R., and Nimmo, J. J. C. (2007). On a direct approach to quasideterminant solutions of a noncommutative KP equation. \textit{Journal of Physics A: Mathematical and Theoretical}, 40(14), 3839.
     
     \bibitem{Gilson2009}
     Gilson, C. R., and Macfarlane, S. R. (2009). Dromion solutions of noncommutative Davey–Stewartson equations. \textit{Journal of Physics A: Mathematical and Theoretical}, 42(23), 235202.
     
     \bibitem{etingof97}
     Etingof, P., Gelfand, I., and Retakh, V. (1997). Factorization of differential operators, quasideterminants, and nonabelian Toda field equations. \textit{arXiv preprint q-alg/9701008}.
     
     \bibitem{riaz2018}
     Riaz, H. W. A., and ul Hassan, M. (2018). Multisoliton solutions of integrable discrete and semi-discrete principal chiral equations. \textit{Communications in Nonlinear Science and Numerical Simulation}, 54, 416-427.
     
   \end{thebibliography}
\end{document}